\documentclass[twocolumn]{aastex62}

\accepted{July 12, 2018}
 \listofchanges 

\shorttitle{Non-thermal emission from stellar bow shocks}
\shortauthors{del Valle \& Pohl}

\begin{document}

\title{Modeling the non-thermal emission from stellar bow shocks}

\correspondingauthor{M. V. del Valle}
\email{mdvalle@uni-potsdam.de}

\author{M. V. del Valle}
\affil{Institute of Physics and Astronomy, University of Potsdam \\
 D-14476 Potsdam, Germany}

\author{M. Pohl}
\affil{Institute of Physics and Astronomy, University of Potsdam \\
 D-14476 Potsdam, Germany}
\affiliation{DESY \\ 
Platanenallee 6, D-15738 Zeuthen, Germany}

\begin{abstract}
Since the detection of non-thermal radio emission from the bow shock of the massive runaway star BD +43$^{\circ}$3654 simple models have predicted high-energy emission, at X and gamma-rays, from these Galactic sources. Observational searches for this emission so far give no conclusive evidence \added{but  a few candidates at gamma rays}. In this work we aim at developing a more sophisticated model for the non-thermal emission from massive runaway star bow shocks. The main goal is to establish whether these systems are efficient non-thermal emitters, even if they are  not strong enough to be yet detected.\deleted{or they are not capable of transforming a significant amount of their power into non-thermal radiation.} For modeling the collision between the stellar wind and the interstellar medium we use 2D hydrodynamic simulations. We then adopt the \replaced{configuration}{flow profile} of the wind and the ambient medium obtained with the simulation as the plasma state for solving the transport of energetic particles injected in the system, and the non-thermal emission they produce. For this \replaced{porpoise}{purpose} we solve a 3D (2 spatial + energy) advection-diffusion equation in the test-particle approximation. We find that a massive runaway star with a powerful wind \replaced{deposits}{converts} \replaced{a fraction of $\sim$ $10^{-5}-10^{-3}$}{0.16-0.4\%}  of \added{the power injected in electrons} \deleted{its total wind kinetic power} \replaced{as}{into} non-thermal emission, mostly produced by inverse Compton scattering \replaced{of relativistic electrons with dust-emitted photons,}{of dust-emitted photons by relativistic electrons, and} secondly by synchrotron radiation. \added{This represents a fraction of $\sim$ $10^{-5}-10^{-4}$ of the wind kinetic power}. Given the better sensibility of current instruments at radio wavelengths theses systems are more prone to be detected at radio through the synchrotron emission they produce rather than at gamma energies.

\end{abstract}


\section{Introduction}
Runaway massive stars are stars with high spatial velocities ($V_{\star} > 30$\,km\,s$^{-1}$) that have been expelled from their formation sites \citep[e.g.,][]{2000ApJ...544L.133H,2011MNRAS.410..190T}. Massive stars have strong winds that interact with the interstellar medium (ISM) as the stars move supersonicaly through it. In this interaction a bow shock is formed, in some cases detectable in the infrared (IR) \citep[e.g.,][]{1988ApJ...329L..93V,2010ApJ...710..549K}. This last emission is reprocessed stellar light by the dust swept by the bow shock. There are of the order of $\sim$ 700 stellar bow shocks cataloged so far \citep{2012A&A...538A.108P,2015A&A...578A..45P, 2016ApJS..227...18K}. 

The bow shock of the massive runaway star BD +43$^{\circ}$3654 was detected at radio wavelengths, and the emission might be synchrotron radiation \citep{2010A&A...517L..10B}. This suggests that a population of high-energy electrons is present in the source, interacting locally with the magnetic field. In the collision between the ISM and the stellar wind a system of two shocks is formed: a forward shock and a reverse shock. This last shock is adiabatic and fast, with velocities of the order of $\sim$ 10$^{3}$\,km\,s$^{-1}$. Hence it is straight forward to think that this reverse shock might accelerate particles up to high-energies through  diffusive shock acceleration (DSA). 

If the electrons that produce the radio non-thermal emission were accelerated in the reverse shock of BD +43$^{\circ}$3654 then they are expected to further interact with the ambient fields: the density and the photons producing high-energy emission via relativistic Bremsstrhalung and inverse Compton (IC) scattering. With this in mind a number of initial models predict non-thermal emission, mainly via IC scattering of IR photons, at X-rays and gamma rays \citep{2012A&A...543A..56D,2014A&A...563A..96D,2016A&A...588A..36P}, see also \citet{2018arXiv180610863D} for a multi-zone model. \citet{2012ApJ...757L...6L} claimed to find the first non-thermal X-ray emission from the bow shock of AE Aurigae, however later it was demonstrated that the emission is not positional coincident with that of AE Aurigae bow shock \citep{2017ApJ...838L..19T}. \added{\citet{2012ApJS..199...31N}}  found an unidentified {\it Fermi} source  locally coincident with the position of the bow shock of the massive star HD 195592, and  \added{\citet{2013A&A...550A.112D}} studied the possibility that this gamma emission was being produced in the bow shock. Although theoretically plausible, in the second {\it Fermi} \replaced{cataloged}{catalog} this source was reclassified as a pulsar \added{\citep{2013ApJS..208...17A}}.

Several searches for high-energy emission from bow shocks of massive runaway stars have followed. At X-rays using both  {\it XMM-Newton} archived observations \citep{2017ApJ...838L..19T,2016ApJ...821...79T} and  dedicated observations \citep{2017MNRAS.471.4452D}, where no non-thermal extended emission was found.  \citet{2017MNRAS.471.4452D} used the derived upper limits at X-rays and those available at radio wavelength  to fit general physical parameters of the sources  with a simple model for the non-thermal emission. They found reasonable fit values for 5 out of the 4 targets of the sample. Also, making energetic assumptions for all the bow shocks listed in the E-BOSS catalog, they concluded that a clear identification of non-thermal X-ray emission from massive runaway bow shocks requires one order of magnitude (or higher) sensitivity improvement with respect to present observatories.

At gamma-ray wavelengths \citet{2014A&A...565A..95S} \replaced{search}{searched} for emission in {\it Fermi} archive data of the 28 bow shocks listed in the E-BOSS catalog \citep{2012A&A...538A.108P}. From the modeled sources only $\zeta$ Oph was detectable, \added{however no emission locally coincident with this source was found in the data; from this it can be concluded that} \deleted{but} the model predictions were \deleted{excluded} \added{overestimated at least} by a factor of $\sim$ 5. For the rest of the sources upper limits were derived in the energy range from 100 MeV to 300 GeV. A study of the same sources was made by the H.E.S.S. collaboration in the energy range between 0.14 and 18 TeV \citep{2017arXiv170502263H}. No associated emission was found but from the resulting upper limits a constraint on the very high energy emission was obtained\replaced{,}{:} it should be less than $0.1$ to $1$\% of the kinetic wind energy.

\added{Recently, \citet{2018arXiv180600614S} presented two runaway stars (Lambda Cephei and LS 2355) whose bow shocks are coincident with two unidentified {\it Fermi} gamma-ray sources from the third {\it Fermi} 3FGL  catalog \citep{2015ApJS..218...23A}. After cross-correlation between different catalogs at distinct wavelengths, the authors found that these bow shocks are the most peculiar objects in the {\it Fermi} position ellipses. Using a simple model for estimating  the IC emission they  fitted the {\it Fermi} data for both sources, obtaining reasonable values for the fitted parameters. This makes these systems promising candidates for gamma-ray bow shocks.}
   
The growing observational base, the progressive interest of the gamma-ray and X-ray community on searching non-thermal emission from stellar sources, together with the new observational upper limits demand now more accurate models of non-thermal emission from runaway star bow shocks. Here we present such a model, aiming to establish new theoretical predictions on non-thermal emission from these sources and also to establish if these systems can be efficient non-thermal emitters. Detailed theoretical work will help to guide the search of these sources at radio and high energies.

In this work we implement a hydrodynamic code to simulate the interaction of the wind of high-mass runaway stars with the ambient medium; then we calculate the non-thermal emission associated with this interaction. Assuming that electrons and protons are accelerated via DSA in the reverse shock we solve the transport of particles and their emissions obtaining emission maps and spectral energy distributions (SEDs). Here we do not focus  in any particular source, that would be addressed in future works.

In the next Section we give a general introduction to the model followed by a more detailed description of the hydrodynamics of the \deleted{the} wind+ISM interaction and our implementation in Sect.\,\ref{sec:hydro}. In Sect.\,\ref{sec:particles} we present  our model for solving the transport of energetic particles. In Sect.\,\ref{sec:results} the obtained results are shown and finally in Sect.\,\ref{sec:discussion} we present a discussion and give our conclusions in Sect.\,\ref{sec:conclussions}. 

\section{Introduction to the Model}\label{sec:intromodel}

As mentioned above, the bow shock of a massive runaway star is formed by the collision of the stellar powerful wind with the incoming ISM, in the star's reference frame. The wind and ISM pressure balance at the contact discontinuity. The characteristic scale of the system is usually taken as the standoff distance $R_0$, given by the balance of the wind and ambient medium ram pressures: 
\begin{equation}\label{eq:R0}
R_0 = \sqrt{ \frac{\dot{M}_{\rm w}V_{\rm w}}{4{\pi}{\rho}_{\rm ISM}v_{\star}^{2}} },
\end{equation}
where $\dot{M}_{\rm w}$ and $V_{\rm w}$ are the wind mass loss rate and velocity, respectively; ${\rho}_{\rm ISM}$ is the ISM density and $v_{\star}$ is the star's velocity. In the {\it instantaneous cooling} approximation $R_0$ would directly give the distance from the star to the apsis of the bow shock, however in a real system this distance might vary, due to thermal conduction and cooling, for example \added{\citep[e.g.,][] {1997RMxAA..33...73R,1998A&A...338..273C,2014MNRAS.444.2754M}}.

In the literature a number of works exists on the collision of two fluids and specifically for modeling the bow shocks of massive runaways \citep[e.g.][]{1996ApJ...459L..31W, 1996ApJ...469..729C,1997RMxAA..33...73R,1998A&A...338..273C,2000ApJ...532..400W,2011ApJ...734L..26V,2014MNRAS.444.2754M,2016MNRAS.459.1146M,2017MNRAS.464.3229M}. A precise description of the phenomenology \replaced{require}{requires} a dynamical treatment implementing numerical simulations. An appropriate treatment of the hydrodynamics of stellar winds  should include both optically thin cooling and thermal conduction \citep[e.g.,][]{1997RMxAA..33...73R,1998A&A...338..273C}. 

   \begin{figure}
   \centering
   \includegraphics[width=0.4\textwidth]{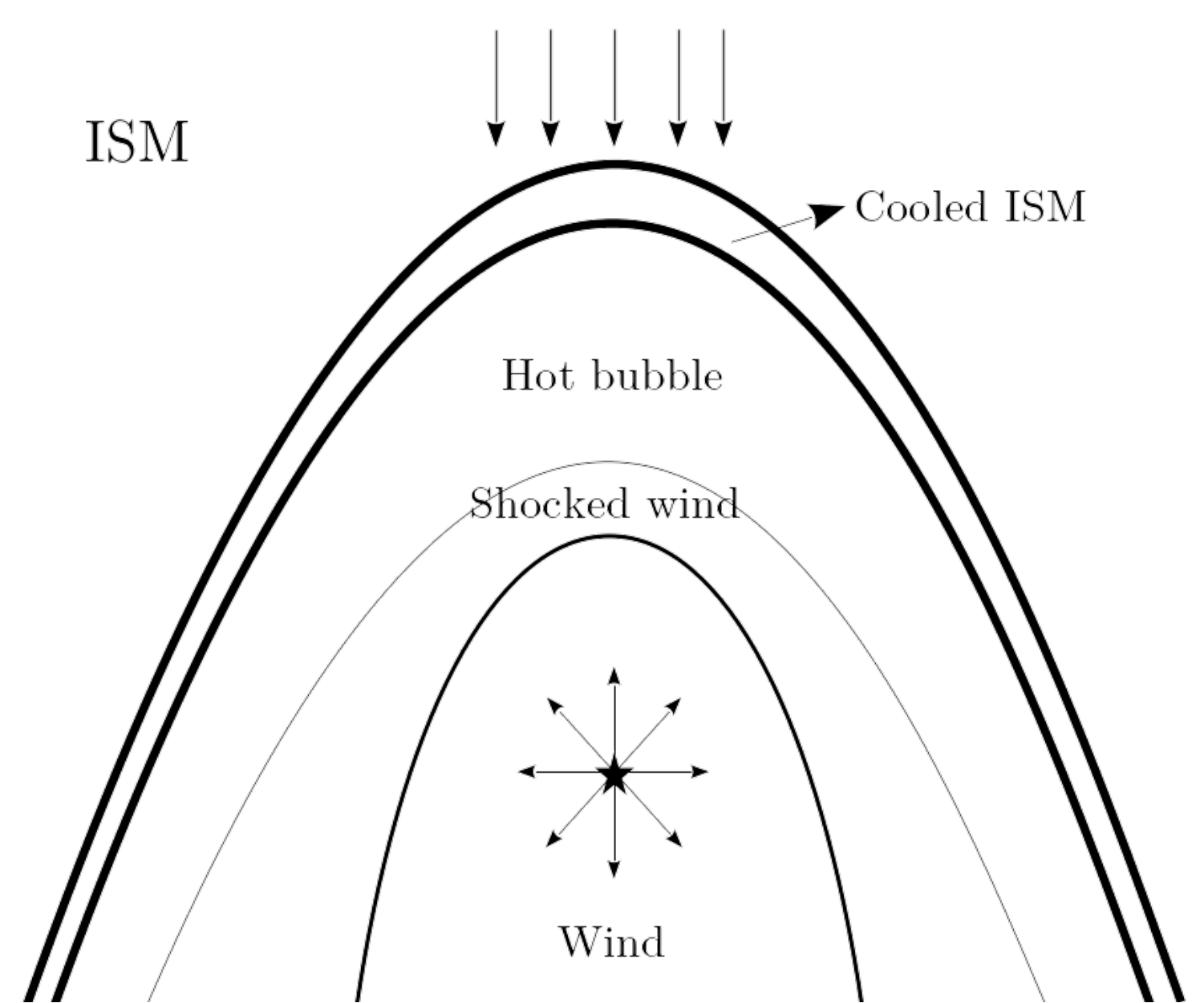}
   \caption{Scheme of a runaway massive star bow shock. Five regions can be distinguished: free flowing stellar wind, the shocked wind, hot shocked ISM, cooled ISM, and the ISM itself. Due to thermal conduction two layers of shocked ISM are formed. A hot and low density layer adjacent to the contact discontinuity and an outer one formed of cooled shocked ambient medium \citep{1998A&A...338..273C,2014MNRAS.444.2754M}.}
              \label{fig:scheme}%
    \end{figure}

After the formation of the bow  shock the system would reach globally a steady state. A general sketch of a bow shock is shown in Figure\,\ref{fig:scheme}. The system is very prone  to suffer many instabilities: Rayleigh-Taylor between the dense cooled layer and the hotter less dense one, an instability arising in shocked layers bounded by thermal pressure on one side and ram pressure on the other \citep[e.g.,][]{1987ApJ...313..820R,1993ApJ...407..207M,1998A&A...338..273C} and Kelvin-Helmholtz due to the velocity shear between the material layers \citep[e.g.,][]{1996ApJ...461..927D}. A complete analysis of these instabilities is made in \citet{1998A&A...338..273C}. 

In this work we use the PLUTO code \citep{2007ApJS..170..228M} to solve the 2D hydrodynamic equations following the set-up by \citet{2014MNRAS.444.2754M} \citep[see also,][]{2016MNRAS.459.1146M,2017MNRAS.464.3229M}. At this stage we do not consider the magnetic field in the simulations. As the system reaches a steady state we use that state as a scenario to solve in it the transport of energetic particles, assumed to be accelerated via DSA in the wind shock. We search for the reverse shock position and \replaced{injected}{inject} there relativistic electrons and protons; using our own code we solve the diffusion-advection equation for the particles in the 2D domain. 

The energetic particles would interact with the magnetic field producing synchrotron emission (only for electrons, proton synchrotron is very inefficient in this case); with the density producing relativistic Bremsstrahlung and $p-p$ inelastic collisions --for electrons and protons, respectively--; and with the radiation fields: the stellar photon field and the stellar-reprocessed dust emission. Only electrons interact efficiently with the radiation fields, via IC scattering.   

Other works that solve the hydrodynamic and magnetohydrodynamic equations together with the transport of high energy particles exist. For example, in \citet{2016A&A...591A..15D} they use a similar approach as the one we use, but here we do solve the spatial diffusion of the particles, which is key in the system we are studying. In \citet{2016ApJ...824L..30P} they solve the hydrodynamics of galactic winds and cosmic-ray diffusion, but they consider this last component as a fluid, without solving the energy dependence of the particles, needed to compute the non-thermal emission; in contrast to this system, the pressure of the energetic particles is negligible in our case. \citet{2016A&A...593A..20B} make a self-consistent treatment of the plasma dynamics, acceleration and transport of cosmic rays in supernova remnants. However their 1D treatment is not appropriate in our problem. 

In the following Sections we describe with more detail the hydrodynamic model and the modeling of the transport of relativistic particles.
 
\section{Hydrodynamic modeling}\label{sec:hydro}

As mentioned previously we use the PLUTO code to solve the 2D hydrodynamic equations following the set-up by \citet{2014MNRAS.444.2754M}. We consider a 2D cylindrical coordinate system with coordinates ($r$, $z$). The system of equations is the following:  

\begin{eqnarray}
	   \frac{\partial \rho}{\partial t}  + 
	   {\bf v}\cdot{\bf \nabla}{\rho} +{\rho}({\bf \nabla}\cdot{\bf v}) & = &  0, \nonumber\\
	   \frac{\partial {\bf v} }{\partial t}  + 
           {\bf v} \cdot {\bf \nabla}{\bf v}  	      + 
           \frac{{\bf \nabla}p}{\rho} 			     & = &  {\bf 0}, \\
	  \frac{\partial p }{\partial t}   + 
	  {\bf v} \cdot {\bf \nabla}p   +
	  \rho c_s^{2}{\bf \nabla} \cdot{\bf v} &  = &	 (\gamma_{\rm g} -1)  
	  \left[{\Phi}(T,\rho) +
	  {\bf \nabla} \cdot {\bf {F}_{\rm c}}\right]; \nonumber
\label{eq:hydro}
\end{eqnarray}

\noindent Here ${\bf v}$, $\rho$ and $p$ are the fluid velocity, its density and pressure, respectively; $\gamma_{\rm g} = 5/3$ is the ratio of specific heats for a monoatomic ideal gas, $c_s$ is the sound speed, $\Phi$ represents the radiative energy gains (heating) and losses (cooling) and ${\bf {F}_{\rm c}}$ is the heat flux due to thermal conduction. 

The total density is $\rho = \mu n m_{\rm H}$ with $n$ the total number density and $\mu = 0.61$, mean molecular weight for a fully ionized medium\footnote{It is shown in \citet{2014MNRAS.444.2754M} that for a massive main sequence star the Str\"{o}mgren sphere is greater than the typical scale of the bow shock. Hence we assume here a fully ionized plasma.}. The temperature $T$ as a function of density and pressure is given by $T =  \mu \frac{ m_{\mathrm{H}} }{ k_{\rm{B}} } \frac{p}{\rho}$, with $k_{\rm{B}}$ the Boltzmann constant.

The radiative term $\Phi$ can be written as  $\Phi(T,\rho)  =  n^{2}_{\mathrm{H}}{\Gamma}_{\alpha}(T)   
		   		 -  n^{2}_{\mathrm{H}}{\Lambda}(T)$, where ${\Gamma}_{\alpha}$ represents the heating and ${\Lambda}$ the optically-thin cooling;  $n_{\rm H}$ is the hydrogen number density, we consider solar abundances. The cooling term includes the cooling of Hydrogen, Helium and metals  \citep[tabulated from][]{2009MNRAS.393...99W}, hydrogen recombination and forbidden lines collisionally exited; the heating term is due to recombination of hydrogen ions. For further details the reader is referred to \citet{2014MNRAS.444.2754M} and references therein.  
		   		 
 The heat flux ${\bf F}_{\rm c} = - \kappa {\bf \nabla}T$ is due to thermal conduction\replaced{, the}{. The} classical heat flux given by Spitzer's coefficient in a fully ionized plasma\replaced{:}{ is} $\kappa = 5.6\times10^{-7} T^{5/2}$ erg\,s$^{-1}$\,cm$^{-1}$.

\subsection{Initial conditions}
We are interested in massive runaway stars with powerful winds, so we consider a typical runaway of mass $M_{\star} = 40 M_{\odot}$, $T_{eff} = 4.25\times10^{4}$\,K, $R_{\star} = 10^{12}$\,cm, $\dot{M}_{\rm w} = 7\times10^{-7}$\,M$_{\odot}$\,yr$^{-1}$, $V_{\rm w} = 2000$\,km\,s$^{-1}$, $V_{\rm star} = 40$\,km\,s$^{-1}$. With these values $R_0 \sim 2.2$\,pc.

We use a rectangular box of size $[0, 24\,{\rm pc}]\times[-18, 8\, {\rm pc}]$ and resolution $(880\times960)$. Initially the box is filled with ISM of density $n_{\rm ISM} = 0.57$ and $T = 8000$ K, and velocity ${\bf v} = - v_{\star} \hat{k}$. The wind is constantly injected in a region $R^{2} = r^{2}+z^{2} < 1\,{\rm pc}$ centered at the origin. Its density is given by $\rho_{\rm w} = \dot{M}_{\rm w}/(4{\pi}v_{\rm w}R^{2})$. We use a tracer (passive scalar) to {\sl color} the wind material. After $\sim$ 16 $t_{\rm cross}$, with $t_{\rm cross} = R/v_{\star}$, the expanding bubble  turns into a steady bow shock \citep{2014MNRAS.444.2754M}.

\subsection{Boundary conditions}
In the initial $r$ boundary, because of the symmetry of the problem, we consider axisymmetric boundary conditions. For the end boundary of both $r$ and $z$ we use {\sl outflow} conditions. Also, we do not allow inflow at the $z$-lower boundary. In the initial $z$ boundary the condition that fresh ISM enters with ${\bf v} = - v_{\star} \hat{k}$ is imposed.  

For solving the dynamic evolution we use a Runge Kutta algorithm of third order, with  linear spatial  reconstruction. These systems are highly prone to instabilities, hence fluxes are \replaced{computing}{computed} using a simple Lax-Friedrichs scheme. This also avoids  2D-effects in the symmetry axis. The parabolic term (thermal conduction) is solved using the Super-Time-Stepping scheme implemented in the code.

\subsection{Results}

\begin{figure*}
\begin{center}
\includegraphics[width=.4\textwidth, angle=270]{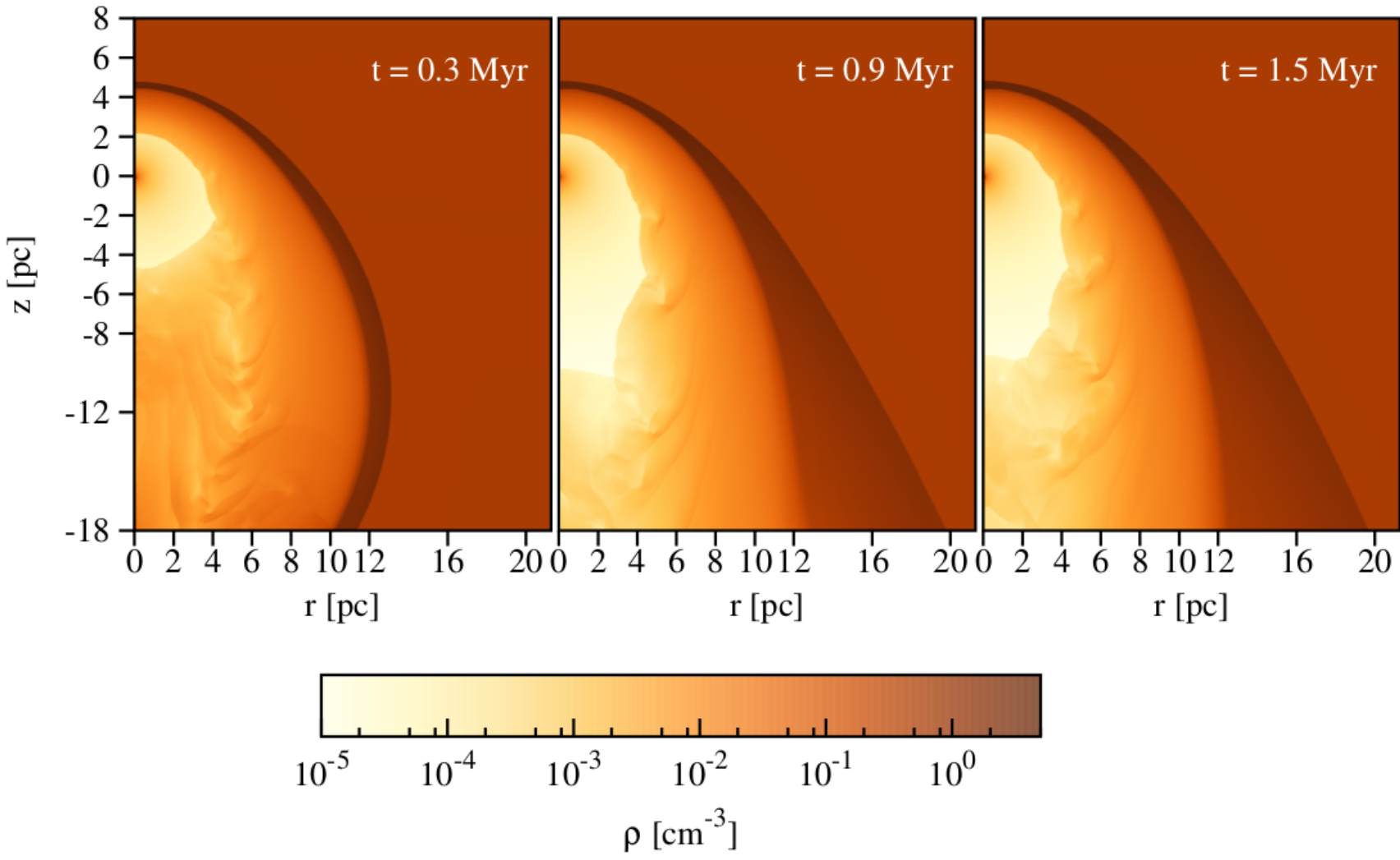}
\caption{Density maps at different computational times for the interaction of a stellar wind with incoming ISM at a velocity $-v_{\star}$. Time evolves from left to right.}
\label{fig:densitymaps}
\end{center}
\end{figure*}

\begin{figure}
\begin{center}
\includegraphics[width=0.7\columnwidth, clip=true,angle=270]{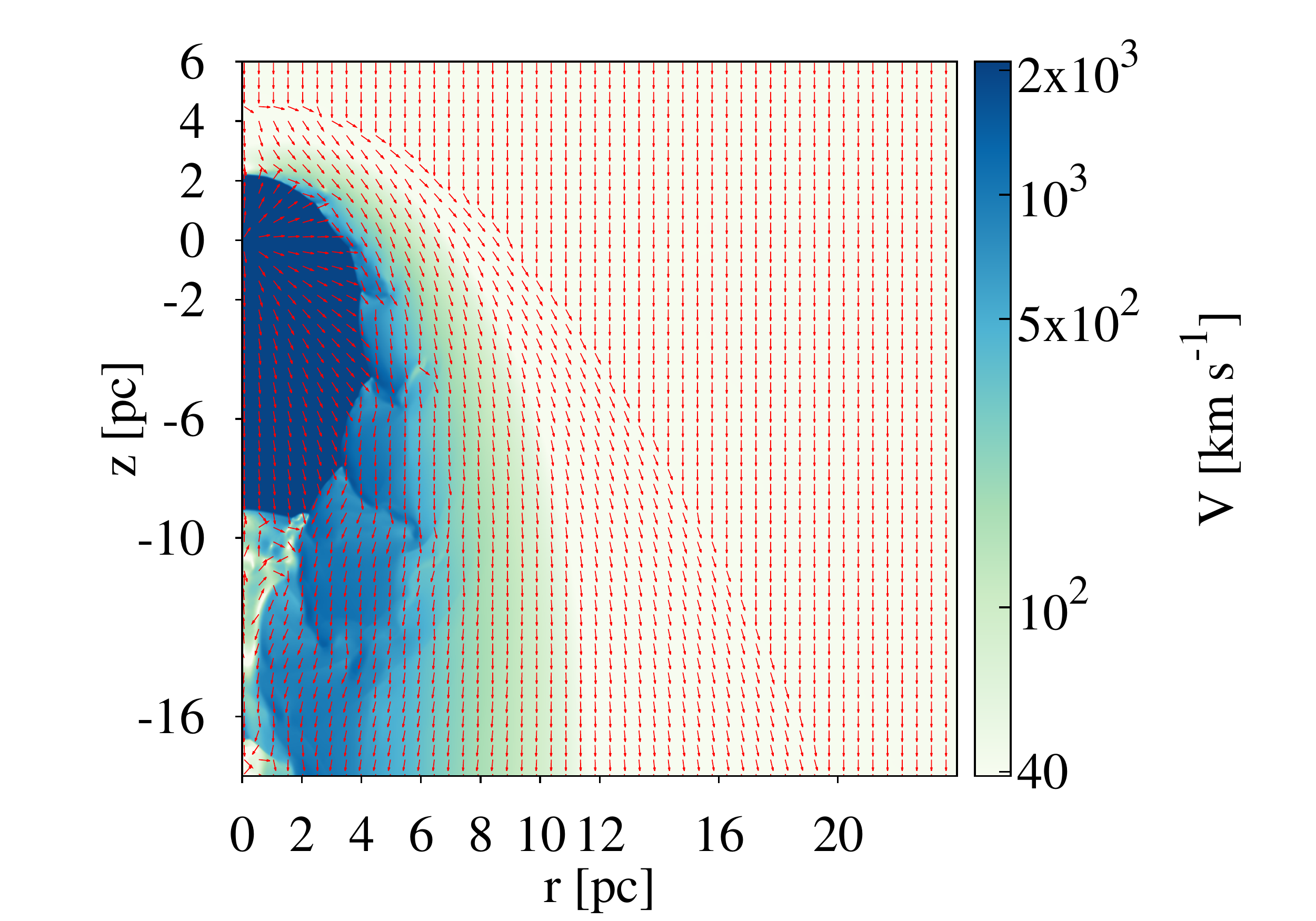}
\includegraphics[width=0.8\columnwidth ,trim= 0cm 0.cm 0cm 0cm, clip=true,angle=270]{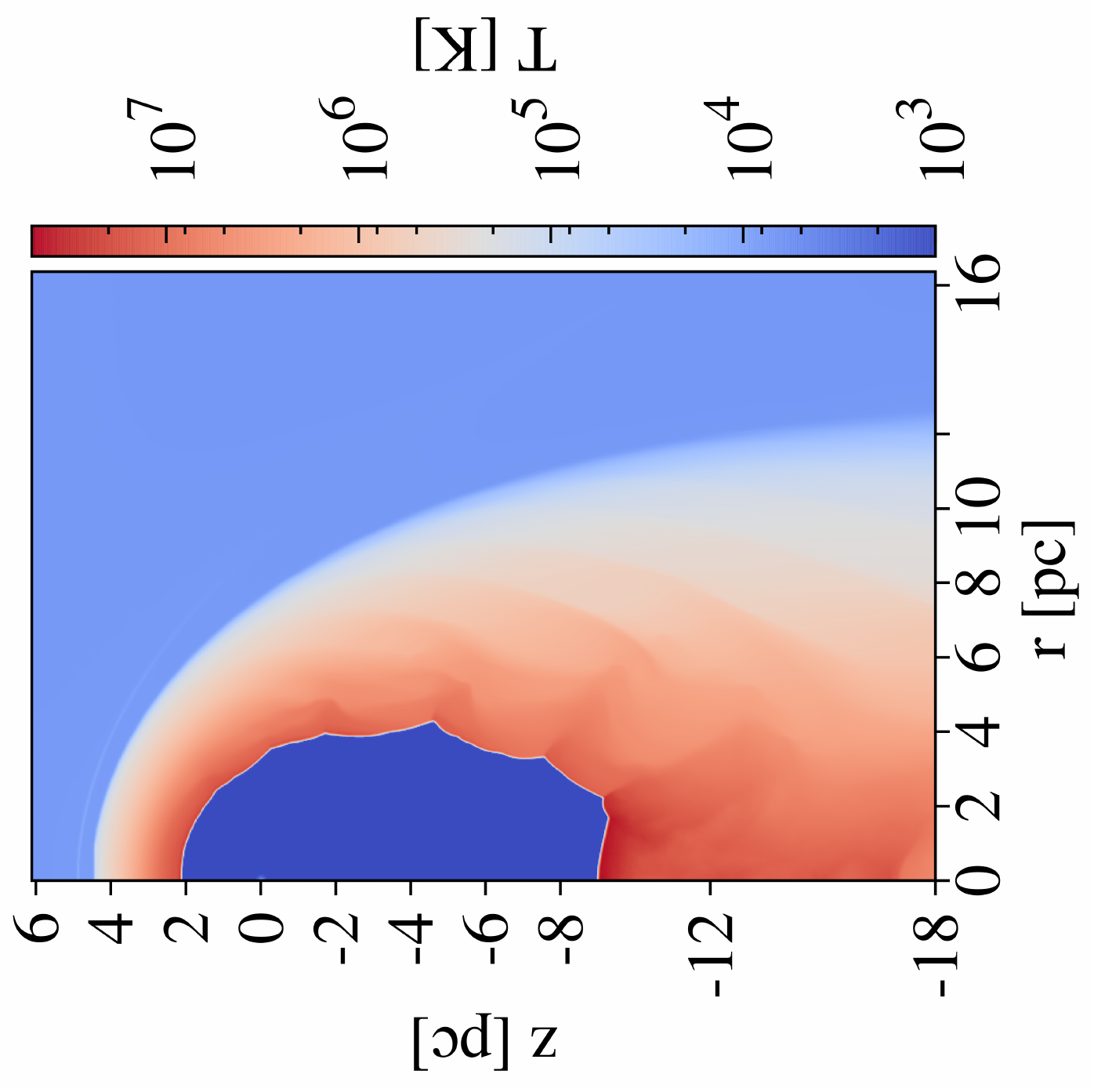}
\caption{Upper plot: Velocity map at $t = 1.5$\,Myr, the colors indicate the value of the velocity module, $\sqrt{v_{\rm r}^{2}+v_{z}^{2}}$, the arrows indicate the direction of the velocity vector ${\bf v}$ in each point. Bottom  plot: Temperature map in the simulation plane, at $t = 1.5$\,Myr.}
\label{fig:velocitymap}
\end{center}
\end{figure}

In Figure\,\ref{fig:densitymaps} we show the evolution of the density in the simulation domain from $t = 0.3$ to $1.5$\,Myr, when the large structure have already reached a steady state. Initially the material expands spherically, the shocked ISM starts to flow surrounding the expanding wind. A thin layer of cooled ISM material starts to form. The structure shows some {\sl fingers} in the hot-cool shocked ISM interface, possibly due to Rayleigh-Taylor instability. At latter times some instabilities in the wind-shocked wind interface appear, possibly due to shear (Kelvin-Helmholtz instability), produced by the different velocities of the two layers (see also Fig.\ref{fig:velocitymap}). Comparing the bow shock shape from $t = 0.9$\,Myr with that at $t = 1.5$\,Myr we can see that there are few  changes in the structure, only in the inner cooling layer due to instabilities; at $t = 1.5$\,Myr the system have already reached a steady state. 

In Figure\,\ref{fig:velocitymap} it is shown (upper plot) a map of the velocity at $t = 1.5$\,Myr. The highest velocities, $\sim$ $10^{3}$\,km\,s$^{-1}$ correspond to the wind, the shocked ISM flows with velocities of the order of hundreds kilometers-per-second or less. In the bottom plot in the same figure it is shown the temperature map, at the same snapshot. The highest temperature corresponds to the shocked wind, with $T \sim 7\times10^{7}$\,K\replaced{, g}{. G}iven the relation $T = 2\times10^{-9} V_{\rm shock}^{2}$\,K for an adiabatic shock of velocity $V_{\rm shock}$, this implies a shock velocity $V_{\rm shock}$ $\sim$ $2\times10^{3}$\,km\,s$^{-1}$ $\sim$ $V_{\rm w}$.

Comparing the cooling time $t_{\rm cool} = \frac{P}{(\gamma_{\rm g} -1 )\Lambda(T)n_{H}^{2}}$ with the dynamical time of an specific layer $t_{\rm dyn} = \Delta z/v$ establishes the \added{adiabatic/}radiative nature of a shock. For the wind shock $T \sim 7\times10^{7}$\,K, $n \sim 10^{-4}$\,cm$^{-3}$, $\Lambda(T)$ $\sim$ $10^{-22}$\,erg\,cm$^{-3}$\,s$^{-1}$ that gives $t_{\rm cool} = 2\times10^{2}$\,Gyr $>>$ $t_{\rm dyn}$ $=$ 0.2\,kyr, with $\Delta z$ $\sim$ 0.1\,pc. For the forward shock $T \sim 10^{5}$\,K, $n \sim 0.5$\,cm$^{-3}$, $\Lambda(T)$ $\sim$ $10^{-21}$\,erg\,cm$^{-3}$\,s$^{-1}$ $t_{\rm cool} = 10$\,kyr, which dominates over the dynamical time that is, for $\Delta z$ $=$ 0.6\,pc, $t_{\rm dyn}$ $=$ 0.2\,Myr. Hence the shock in the wind is adiabatic, and the shock in the ISM is radiative. 

\begin{figure}
\begin{center}
\resizebox{1.\columnwidth}{!}{\includegraphics[scale=.3,clip=true,angle=270]{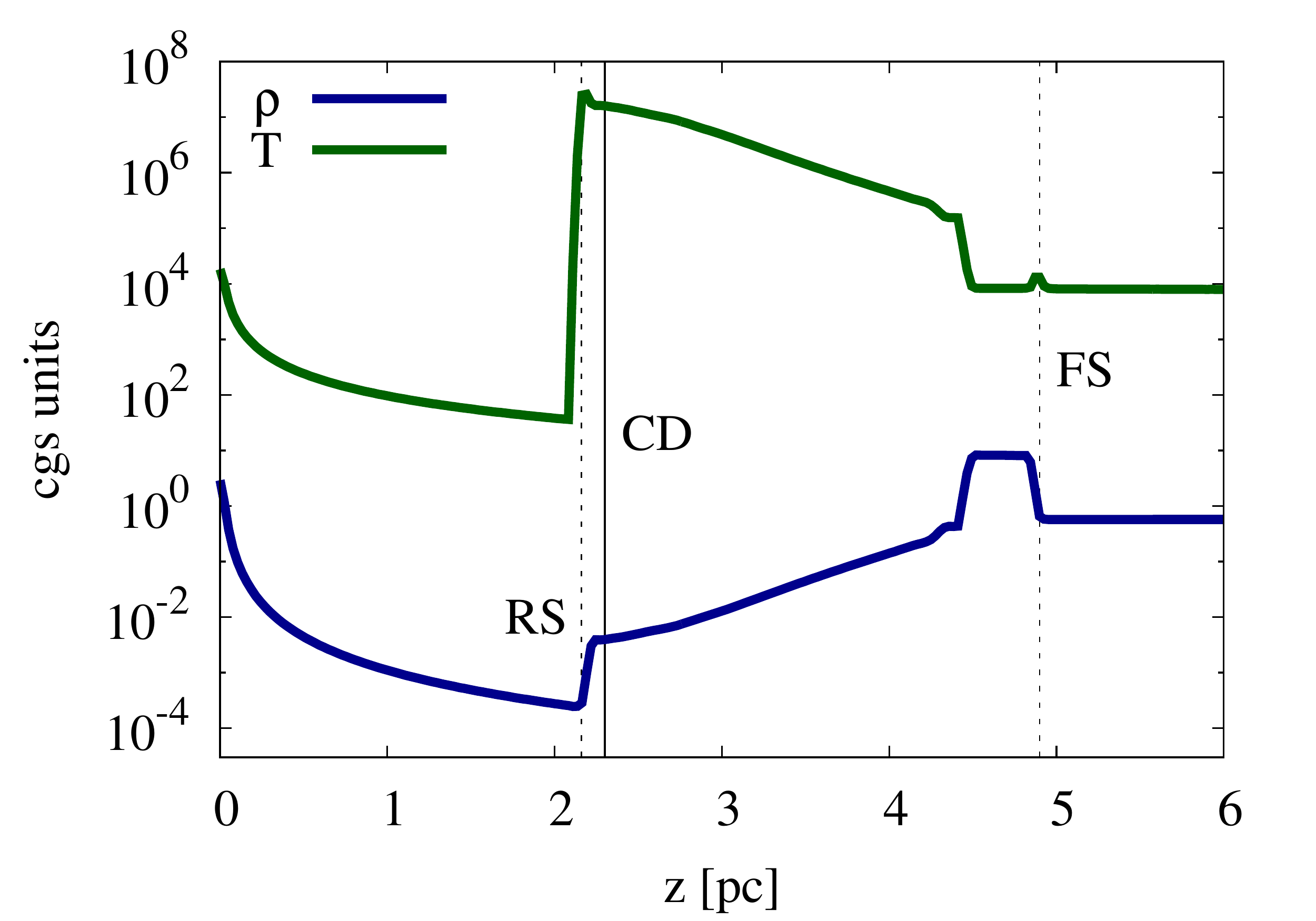}}
\caption{Density and temperature profiles for $r = 0$ as a function of $z$, the units are cm$^{-3}$ and $K$, respectively. The reverse shock and forward shock positions are marked with dashed lines.}
\label{fig:profile}
\end{center}
\end{figure}

Profiles of density and temperature for $r = 0$, $z \geq 0$ are shown in Figure\,\ref{fig:profile}, at $t = 1.5$\,Myr. The density decreases radially as expected from ${\rho}_{\rm w} \propto R^{-2}$, a density jump occurs at $z \sim 2.16$\,pc that coincides with a jump in the temperature: this is the wind or reverse shock. The contact discontinuity, marked in the figure with a solid vertical line, is located at $z \sim 2.3$\,pc, a bit further than the position predicted  theoretically, i.e. $z \sim R_0$, an effect expected by thermal conduction \citep[e.g.,][]{1998A&A...338..273C}. The density increases slowly after the jump\replaced{, t}{. T}his increase of mass in the intermediate density layer is caused by thermal conduction \citep{1988ApJ...329L..93V}. At  $z \sim 4.9$\,pc another jump in density is encountered, again in company of a jump in temperature, this is the forward shock. The dense layer of the bow shock is in thermal equilibrium with the ISM.

It is worth mentioning that this profile differs from the more \replaced{"stepy"}{sharply structured} typical reverse shock-contact discontinuity-forward shock profile in which the regions of different materials are well delimited. The presence of thermal conduction produces an intermediate density layer. The temperature of the shocked ambient gas is much lower than that of the shocked wind, this causes a flow of energy outwards; in turn a inward flow of matter from the dense layer into the shocked wind region occurs \citep[][]{1998A&A...338..273C}.  

\citet{2017MNRAS.464.3229M} demonstrated that the presence of a ISM magnetic field does not change the global shape of the bow shock, but it  modifies the thermal conduction and hence the hot bubble size. The magnetohydrodynamic treatment will be addressed in a future work.  Analyzing deeply the hydrodynamic of the system is not the goal of this work, many previous works --mentioned in Sect.\ref{sec:intromodel}-- have done this extensively and the readers are \replaced{refereed}{referred} to them for further \replaced{inquires}{inquiries} in the subject. 

\section{Transport of high-energy particles modeling}\label{sec:particles}

We solve the transport of electrons and protons in the bow shock of the massive star described in the previous section, using  the solution of the HD simulations at $t = 1.5$\,Myr. We use the same cylindrical coordinate system $(r,\;z)$. The diffusion-advection equation  for relativistic protons and electrons that follows $N(t,\,E,\,\vec{r})$ $\equiv$ number of particles $/$ unit energy $\times$ unit volume, is: 

\begin{eqnarray}\label{eq:transport}
 & \frac{\partial N(t,\,E,\,\vec{r})}{\partial t}
=  \nabla \left(D(t,\,E,\,\vec{r})\nabla N(t,\,E,\,\vec{r})\right) - \nonumber\\
 & \nabla \left( \vec{v}(t,\,\vec{r})N(t,\,E,\,\vec{r})\right)  
   - \frac{\partial}{\partial E} \left(P(t,\,E,\,\vec{r})\,N(t,\,E,\,\vec{r})  \right)  \nonumber\\ 
  & + \,Q(t,\,E,\,\vec{r}),
\end{eqnarray}

\noindent where the first term represents the diffusion in space with diffusion coefficient $D(t,\,E,\,\vec{r})$, followed by the advection term with $\vec{v}(t,\,\vec{r})$ the fluid velocity; the third term corresponds to radiative losses where $P(t,\,E,\,\vec{r})$ is the energy loss rate for a particle with energy $E$. Finally $Q(t,\,E,\,\vec{r})$ is the injection function, i.e. number of injected particles $/$ unit energy $\times$ unit volume $\times$ unit time. 

We solve Eq.\,\ref{eq:transport} in a 3D grid $\equiv$ $(E,\,r,\,z)$ using our own modular code \citep[see,][]{2015MNRAS.448..207D,2017arXiv171106250D}. In what follows we describe each of the terms and the model details.

\subsection{Injection}\label{sec:injection}

As argued above the particles are thought to be accelerated in the reverse shock. Here we do not simulate directly the acceleration of the relativistic particles, instead we assume that particles are accelerated at a rate $t_{\rm acc} = {\eta} {r_{\rm L}}/{c}$ \citep[e.g.,][]{1990cup..book.....G}. Here $r_{\rm L}$ is the Larmor radius of a particle of energy $E$, i.e $r_{\rm L} = E/e\,B$ and $B$ is the magnetic field in the acceleration region. $\eta$  is a phenomenological parameter related to the efficiency of the acceleration process, which can be approximated by $\eta \sim {20}/{3}\left({c}/{V_{\rm shock}} \right)^{2}$ \citep{1983RPPh...46..973D}, for a non-relativistic diffusive shock acceleration, in a plane shock in the test particle approximation.

We inject continuously a population of relativistic ($E > mc^{2}$) electrons and protons at the reverse shock position $(r_{\rm rs}, z_{\rm rs})$ (see below). This shock is strong everywhere, however the density in the regions of positive $z$ is greater than in the negative region (this is simply because these points are further away from the star) while the wind velocity remains constant. We expect more particles to be injected in the denser regions, hence the injection function scales as $\propto$ $\rho(r,z)$. The particles have a power-law distribution in energy of index $\alpha = 2$, as expected from a DSA mechanism. Then the injection function reads:
\begin{equation}
Q(t,\,E,\,r,\,z) = Q_0\,E^{-\alpha}\,\rho(r,z)/\rho_0\,\,\delta^{2}\left(\vec{r} - \vec{r}_{\rm rs}\right).
\end{equation}

\noindent $\rho_0$ is a reference density value \deleted{taken that} \added{considered} at the apsis of the wind shock; $Q_0$  is a  normalization factor which depends on the power available in the system for particle acceleration.  

The source power for accelerating the particles is the kinetic power of the wind $L_{\rm w} = 0.5\dot{M}V_{\rm w}^{2}$. A fraction $\xi$ of this  kinetic power is transferred to the particles in the acceleration process. Then the power in relativistic particles is  $L_{\rm rel} = \xi L_{\rm w}$\deleted{, equally divided between electrons and protons}. We use a rather modest value of  $ \xi = 0.05$,  for the system considered here \replaced{$L_{e,\,p} \sim 2.5$}{$L_{\rm rel} \sim 4.4$}$\times10^{34}$\,erg\,s$^{-1}$. \added{The proton-to-electron flux ratio $a$, for an acceleration process that leads to a power-law in momentum (the same for electrons and protons) is calculated in \citet{1993A&A...270...91P}. In the later work the authors assumed that the same number rate of electrons and protons are accelerated from the same initial energy for both species. For a power-law index $\alpha \neq$ 2, $a = \left(m_p/m_e\right)^{(3-\alpha)/2}$. In the case  $\alpha = 2$, $a$ depends very weakly on the particles maximum energies, and is of the order of $ \sim 40$. Starting from $a \sim 40$, the value of $a = 100$ observed in cosmic rays can be easily explained from propagation effects. In this work we assume $a = 40$ and in Sect.\,\ref{sec:discussion} we discuss the implications of adopting others values for $\xi$ and $a$.}

For obtaining the position $(r_{\rm rs}, z_{\rm rs})$ of the reverse shock we search for a jump in the temperature function $T(r,\,z)$  (shown in the bottom plot of Figure\,\ref{fig:velocitymap}), in the wind material.

\subsection{Diffusion}

Stellar winds are very turbulent systems, in particular the system we are studying in which the wind collides with the incoming ISM (see Sect.\,\ref{sec:hydro}). In such scenario slow particle diffusion is expected, as in the case of the sun where a particle of $E \sim 100$\,MeV in the solar wind has a mean free path of $\lambda \sim 1$\,AU, that gives a diffusion coefficient $D \approx 10^{23}$\,erg\,s$^{-1}$.  

We assume the diffusion coefficient to depend only on the particles energy, i.e. $D(t,\,E,\,\vec{r}) \equiv D(E)$. Close to the shock the diffusion is in the Bohm regime, at certain scale a transition occurs between this slow Bohm diffusion to the fast diffusion estimated  in the  Galaxy \citep[e.g.,][]{2012A&A...541A.153T}. The characteristic scale of the system we are studying here, given by Eq.\,\ref{eq:R0}, is of the order of parsecs, hence \replaced{is}{the} more convenient \replaced{to assume}{assumption of} a {\sl Galactic-like} diffusion coefficient:
\begin{equation}\label{eq:diff}
  D(E) = D_{10\,{\rm GeV}} \left(\frac{E}{10\,{\rm GeV}}\right)^{\delta}\,{\rm cm}^{2}\,{\rm s}^{-1}.
\end{equation}
\noindent Here $D_{10\,{\rm GeV}}$ is the value of the diffusion coefficient at $E = 10\,$GeV and $\delta$ is a power-law index varying in the interval 0.3 and 0.6 depending on the power-law spectrum of the
turbulence of the magnetic field. Typical values for the Galaxy are $D_{10\,{\rm GeV}} = 10^{28}$\,cm$^{2}$\,s$^{-1}$ and $\delta = 0.5$ \citep[e.g.,][]{1990acr..book.....B}. As discussed above in this system values much lower than this  are expected due to the presence of turbulence.

In this work we use $\delta = 0.5$ and two values for  $D_{10\,{\rm GeV}}$: $10^{25}$\,cm$^{2}$\,s$^{-1}$ for the slow case, and $10^{27}$\,cm$^{2}$\,s$^{-1}$ for a fast diffusion situation. We can estimate a characteristic timescale for diffusion $t_{\rm diff}$ considering the typical spatial scale of the problem $R_0$, also this is approximately the minimum distance between the injection position and the bow shock itself (the dense cooled ambient matter), 
\begin{equation}
t_{\rm diff} \sim \frac{R_0^2}{D_{10\,{\rm GeV}}}.
\end{equation} 
\noindent Then, $t_{\rm diff}$ $\sim$ $1.5$ and $150$\,kyr for fast and slow diffusion, respectively (see Figure\,\ref{fig:tscale}). 

\subsection{Advection}\label{sec:advection}

The velocity field responsible for the advection of particles is shown in the upper plot of Figure\,\ref{fig:velocitymap}. As the system is in steady state, $\vec{v}$ does not depend on time. We can distinguish here between wind advection and ISM advection.

We can estimate a characteristic timescale $t_{\rm adv}$, as done above for the diffusion, for the wind:
\begin{equation}
t_{\rm adv, \,w} \sim \frac{R_0}{v}.
\end{equation} 
\noindent The velocity is $v \leq 2\times10^{3}$\,km\,s$^{-1}$, then $t_{\rm adv, \,w} \geq 1$\,kyr.

The vertical advection produced by the ISM is relevant almost everywhere, for  $v = 40$\,km\,s$^{-1}$, then $t_{\rm adv, \,ISM} \sim 54$\,kyr. These time scales are plotted in Figure\,\ref{fig:tscale}. It is clear that advection dominates the transport in the case of slow diffusion. A particle injected at $z \sim -9$\,pc would reach the bottom boundary in $\sim$ 200\,kyr. In that time, for the slow regime\deleted{n}, a 10\,GeV particle would radially diffuse approximately 2.7\,pc before it reaches the bottom boundary, and a TeV particle 8.5\,pc. 

Towards the $+z$ direction the situation is more complicated because the advection in the inner regions of the bow shock is not vertical. After being injected the particles are advected in the shocked wind, with $v \sim v_w /4.0$. The particles reach out a distance $\approx$ $D/v$ $=$ 0.08 - 0.24\,pc for $E = 10$\,GeV - 1\,TeV, respectively. In the case of fast diffusion these distances are two orders of magnitude higher.  

\subsection{Non-thermal losses}

The third term in Eq.\,(\ref{eq:transport}) accounts for the relevant non-thermal \replaced{loses}{losses} that particles suffer after their injection in the system. For electrons the non-thermal processes considered are: relativistic Bremssthalung, synchrotron, and IC scattering with the stellar and reprocessed stellar photons (dust emission). For protons the only energy losses considered are due to $p-p$ inelastic collisions. All the target fields: magnetic field, density and radiation fields are inhomogeneous. The density field is directly taken from the simulations, below we describe how we construct the rest of the fields.

\subsubsection{Magnetic field}

We reconstruct the magnetic field from the stars's magnetic field $B_{\star}$, the ISM magnetic field $B_{\rm ISM}$ and density compression. We \replaced{assume}{assumed} no \replaced{prefer}{preferred} direction for the field, which is \replaced{assume}{assumed} to be randomly distributed in all the domain. We consider four regions: the stellar wind region, the shocked wind, the shocked ISM and the ISM itself. For the wind region we use the approach made in  \citet{1982ApJ...253..188V}, assuming flux conservation they obtained a field  which decreases $\propto$ $R^{-1} = \sqrt{r^2+z^2}$:
\begin{equation}\label{eq:magwind}
B_{\rm wind} = B_{\star}
 \left[ 1 + \left( \frac{V_{\rm w}}{V_{\rm rot}} \right)^{2} \right]^{-1/2}\left( \frac{R_{\star}}{R} \right) 
\left[ 1 + \left( \frac{R_{\star}V_{\rm w}}{R\,V_{\rm rot}} \right)^{2} \right]^{1/2},
\end{equation}
\noindent $V_{\rm rot}$ is the rotational velocity (we use a typical value of 100\,km\,s$^{-1}$). In the reverse shock the magnetic field is allowed to compress by a similar factor as the density. Beyond the discontinuity\footnote{The position of this discontinuity is computed using the tracer values at $t = 1.5$\,Myr.} between the wind and ambient material at the coordinates $(r_{\rm dis},\,z_{\rm dis})$ the magnetic field is \replaced{assume}{assumed} to be that of the ISM rescaled with the density field at each point.  

Hence, $B(r,\,z)$ reads:
\begin{eqnarray}\label{eq:magneticfield}
& B(r,\,z) = \nonumber\\ 
&             \left\{ \begin{array}{lcc}
              B_{\rm wind} &   {\rm if}  & (r,\,z) \le (r_{\rm rs},\,z_{\rm rs})  \\

             \\ B(r_{\rm rs},\,z_{\rm rs})\times \mathcal{F}_1  &   {\rm if}  & (r_{\rm rs},\,z_{\rm rs}) \le (r,\,z) \le (r_{\rm dis},\,z_{\rm dis}) \\ 
             \\ B_{\rm ISM} \times \mathcal{F}_2 &  {\rm if}  &   (r_{\rm dis},\,z_{\rm dis}) \le (r,\,z) \le (r_{\rm ISM},\,z_{\rm ISM})\\
             \\ B_{\rm ISM} &  {\rm if}  & (r,\,z) \geq (r_{\rm ISM},\,z_{\rm ISM})
             \end{array}
   \right.
\end{eqnarray}

\begin{figure}
\begin{center}
\includegraphics[scale=.37,trim=0cm 0cm 0cm 0cm, clip=true,angle=270]{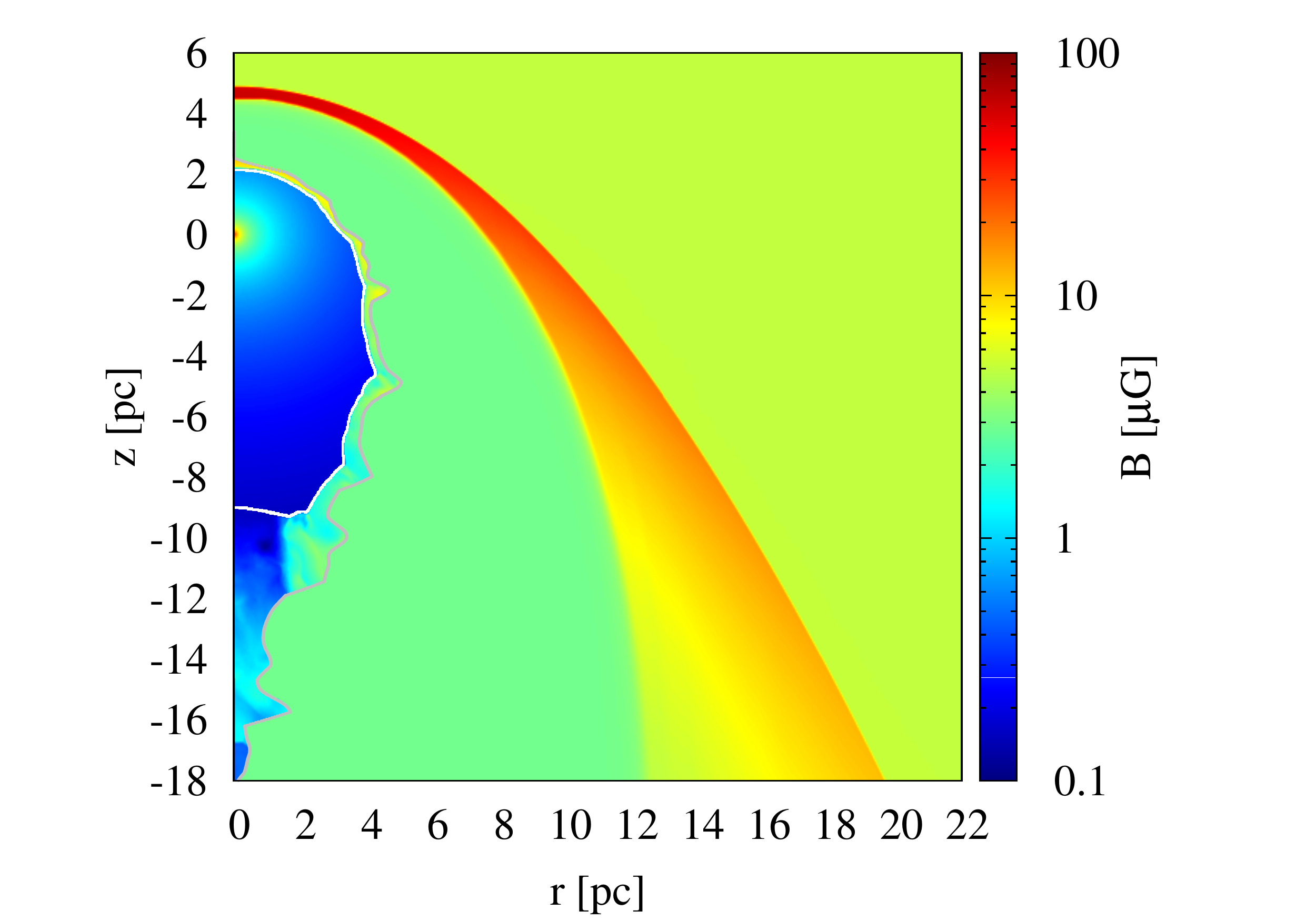}
\caption{Magnetic field map in the computational domain, reconstructed from Eq.(\ref{eq:magneticfield}), $B_{\rm ISM}$ and the density map. The white curve shows the position of the wind shock, where the particles are injected into the domain; the grey curve shows the material discontinuity.}
\label{fig:magneticfield}
\end{center}
\end{figure}

\noindent Where $\mathcal{F}_{1,\,2} = \sqrt{2(\mathcal{K}^{2} _{1,\,2}-1)/3+1}$, with $\mathcal{K}_1 = \rho(r_{\rm rs},\,z_{\rm rs})/{\rho(r,\,z)}$, and $\mathcal{K}_2 = \rho(r_{\rm ISM},\,z_{\rm ISM})/{\rho(r,\,z)}$. \replaced{These last factors}{The factors $\mathcal{F}_{1,\,2}$} account for the shock compression effect in the random field; for a strong shock $\mathcal{K} = 4$ and  $\mathcal{F} = \sqrt{11}$. Here we use $B_{\star} \sim 100$\,G \citep{2012SSRv..166..145W}, however we consider a greater value in Sect.\,\ref{sec:results}. For the ambient medium we use $B_{\rm ISM} \sim 5\,{\mu}$G. The Figure\,\ref{fig:magneticfield} shows the map of the magnetic field in the computational domain; superimposed in white is plotted the position of the reverse shock: the injection position and in grey the material discontinuity between the shocked wind/ISM.

\graphicspath{./}

\subsubsection{Target radiation fields}

As \replaced{stayed}{stated} before the target radiation fields are those from the star and from the bow shock itself. The stellar photon field is \replaced{assume}{assumed} to be that of a black body at $T_{\rm eff}$, decaying as $R^{-2}$ away from the star. For the IC emission calculations we assume the field to be monoenergetic, with $E_{\rm ph} = k_{\rm B}T_{\rm eff}$. 

Computing the radiation field for the reprocessed emission is more complicated because it requires adopting  a dust model. The emission of the bow shock, mainly at IR, is produced by dust heated by starlight\footnote{The produced emission by collisionally heated dust grains is subdominant in these systems \citep{2014MNRAS.444.2754M}.}. Here we adopt a thermal approximation for the dust emission, with grains in thermal equilibrium\replaced{; this}{. This} treatment is appropriate given that the observed fluxes of the bow shocks are in the mid (MIR) to far IR (FIR). 

In order to calculate the equilibrium temperature $T_{\rm gr}$ of the dust grains we equate the absorbed energy with the emitted one. In equilibrium the absorbed energy by the dust should be the same energy it radiates \citep{2005ism..book.....L,2011piim.book.....D}:
\begin{equation}\label{eq:balance}
\int_{0}^{\infty} I_{\nu} \pi a^2 Q_{\rm abs} (\nu) \,{\rm d}\nu = \int_{0}^{\infty} 4 \pi a^2 Q_{\rm em}(\nu) \pi B_{\nu}(T_{\rm gr})\, {\rm d}\nu.
\end{equation}
\noindent Here we use \replaced{an}{a} spherical  dust grain of radius $a$.  The left-hand-side is the frequency integration of the incident flux from the star, that scales with distance as $R^{-2}$, multiplied by the absorption efficiency and the grain cross-section. The right-hand-side is the integration over frequency of the surface of the grain times the emitted spectrum. The dust emissivity is a modified black body at $T = T_{\rm gr}$, this is  $B_{\nu}(T_{\rm gr})$ multiplied by an emissivity function $Q_{\rm em}(\nu)$. The emissivity function is a power law in frequency, we use a standard model with $Q_{\rm em}$ $\propto$ ${\nu}^{2}$. 

For estimating the temperature above  we use the so-called Plank-averaged absorption in the ultraviolet (UV) $\langle Q_{\rm abs} \rangle$ and emission efficiencies $\langle Q_{\rm em} \rangle$ in the IR \citep{2011piim.book.....D}. In the UV the absorption efficiency can be approximated by unity, this is valid as long as the \replaced{sizes of the  grain}{grain sizes} are of the order of the UV photons wavelengths\footnote{This is the case for the relative large dust grains responsible for the IR radiation detected in massive runaway stars bow shocks.}(i.e., $0.01 \leq \lambda \leq 0.4$\,$\mu$m). The grain temperature depends on the position, and is given by:
\begin{equation}\label{eq:temperature}
T_{\rm gr} = \left(\frac{R_{\star}}{\sqrt{r^2+z^2}}\right)^{1/3}\frac{T_{\star}^{2/3}}{(4{\pi}\langle Q_0 \rangle )^{1/6}a_{\mu{\rm m}}^{1/3}}.
\end{equation}
\noindent We assume no dust in the stellar wind region. Dust grains exhibit a distribution of sizes, believed to be a power law in $a$\replaced{, here for simplicity}{. For the sake of simplicity,} we consider that all grains have the same radius. The dust temperature is of the order of $T_{\rm gr} \sim 100$\,K, which is consistent with bow shocks being detected in the IR  at $\lambda \sim 22$\,${\mu}$m \citep[e.g.,][]{2012A&A...538A.108P}, because the maximum of the dust radiation occurs at $\lambda_{\rm max} \sim 2\times 10^{3} / T_{\rm gr}\,{\mu}{\rm m}$. 

The emissivity depends on the amount of dust, i.e. it scales with density, and temperature at each point. The energy loss  by IR emission for one dust grain, i.e. the power emitted, is given by: $P_{\rm gr} = 4{\pi}\langle Q_{\rm em} \rangle_{T_{\rm gr}}\,{\sigma}T_{\rm gr}^{4}$ \citep{2011piim.book.....D}\replaced{, then}{. Then} for a number of grains per unit volume $n_{\rm gr}$, the total power per unit volume  is $P = n_{\rm gr}P_{\rm gr}$. For computing $n_{\rm gr}$ we  assume a typical gas-to-dust density ratio of 100 and we estimate the mass of each dust grain as $m_{\rm gr} = 4 {\pi} a^{3}\rho_{\rm gr}$, with  $\rho_{\rm gr} \sim 2$\,gr\,cm$^{-3}$ \citep{2007ApJ...657..810D}. The resulting expression is (all units are in cgs):

\begin{equation}\label{eq:power}
P(r,\,z) = \chi \frac{3}{200} \,\langle Q_0 \rangle \,\sigma T_{\rm gr}^{6}(r,\,z) \,\rho (r,\,z)\,\,\,\,{\rm erg}\,{\rm s}^{-1}\,{\rm cm}^{-3}. 
\end{equation}

\noindent Here $\chi$ is a factor such that the luminosity from dust in the region does not exceed the star luminosity, i.e. $4\pi \sigma T_{\star}^{4} R_{\star}^2$. 

\begin{figure}
\begin{center}
\includegraphics[scale=.4,trim=0cm 2cm 0cm 0cm, clip=true,angle=270]{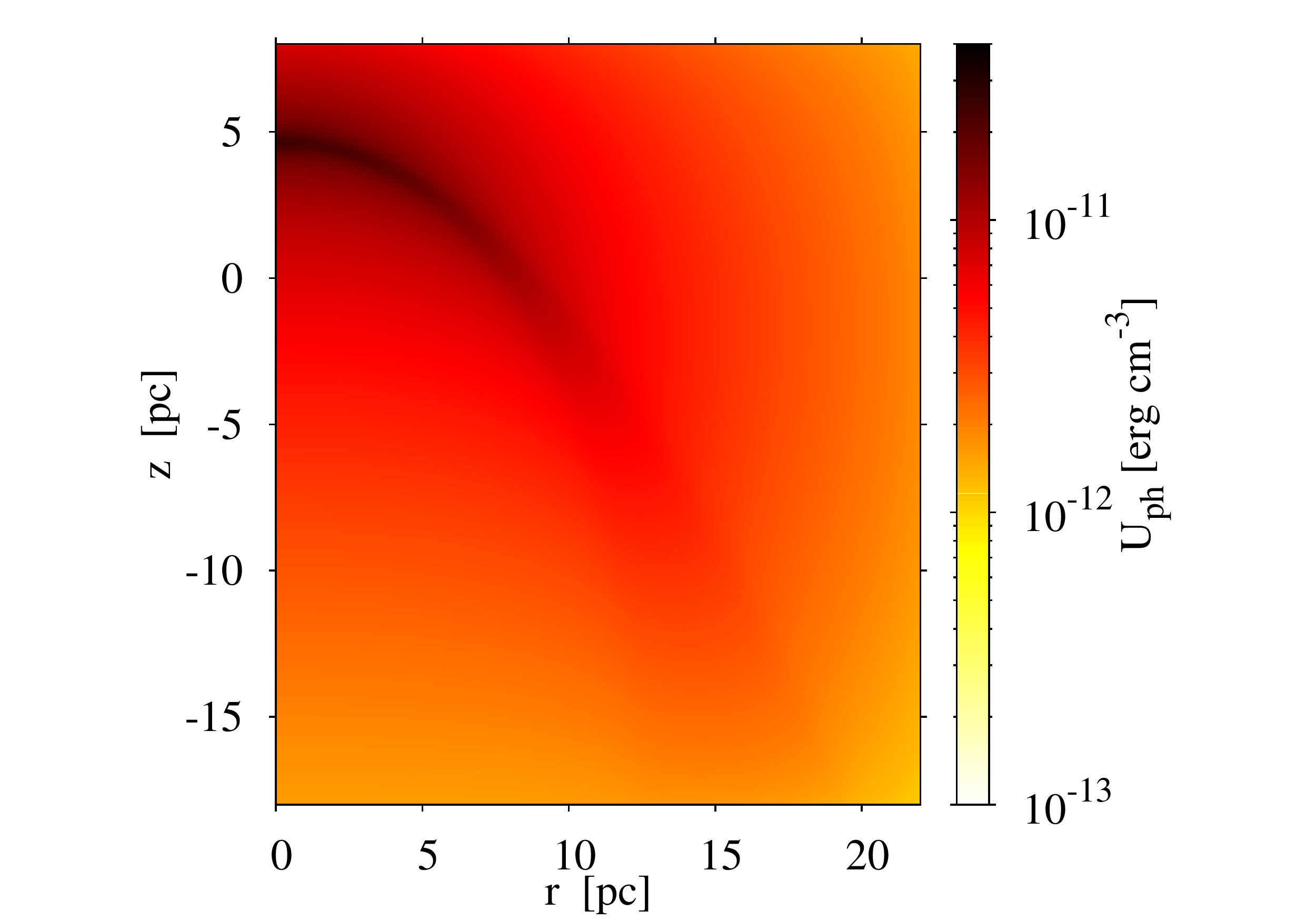}\\
\caption{Infrared photon target energy density in the computational domain.}
\label{fig:energydensity}
\end{center}
\end{figure}

For obtaining the energy density of the photon field in each point we compute $U_{\rm ph}(r,\,z) = \int P / (4{\pi}\,c\,d^{2})\,{\rm d}V$, with $d\equiv d(x,y,z)$ the distance of each point \deleted{in space} to the emitting source \added{and $P$ is} given by  expression \ref{eq:power}. The energy density maps of the target IR photon field is shown in Figure\,\ref{fig:energydensity} for grain size $a_{\mu{\rm m}} =$ 0.01\replaced{; even}{. Even} though $P$ does not depend explicitly on  the size of the grain it depends strongly on the dust temperature. In a real source the grains responsible for the IR radiation have a size distribution, however the grain size distribution is a power law with index smaller than $-3$, then it is more probable to encounter smaller dust grains. Note that the grains responsible for the stellar photons absorption can not be smaller than $a_{\mu {\rm m}} = 0.01$. The  IR photon field  in the IC calculations is also assumed as monoenergetic, with $E_{\rm ph} = k_{\rm B}\langle T_{gr}\rangle$, where  $\langle T_{gr}\rangle$ is the mean grain temperature in the computational region.

\graphicspath{./}

\subsection{Maximum particle energies}\label{sec:maxener}

The maximum energy that particles achieve in a DSA process depends on many factors\replaced{; the}{. Its} estimation \deleted{of it} is not straight forward given that the mechanism is non-linear. However we can make an order of magnitude estimation by comparing the gain rate per energy with the losses in the acceleration region (only in the case of electrons, in the case of protons their energy \replaced{loses}{losses} are not limiting their acceleration) or by the limit imposed by the \added{size of the} acceleration region (a constraint valid for both electrons and protons). 

For estimating $E_{\rm max}$ from the losses we equate $t_{\rm acc} = {\rm min}( t_{\rm synchr}(r_{\rm rs},\,z_{\rm rs}); t_{\rm IC}(r_{\rm rs},\,z_{\rm rs}))$. \deleted{In the case we use the size of the acceleration region as a constraint we are assuming that the acceleration takes place in the reverse shock, given the characteristics} \added{Given the physical size} of the system \added{(see Fig.\,\ref{fig:densitymaps} and Fig.\,\ref{fig:profile})} the acceleration process then should proceed in a region of size the order of $l$ $\sim$ $1$\,pc. Imposing the condition that the precursor size should be smaller than 1\,pc and assuming Bohm diffusion for the acceleration,  we obtain the maximum energy, i.e. $E < 3\,e\,B|_{\rm shock}V_{\rm shock}l/c$, we use 10\% of this value. Both methods for obtaining the maximum energy are sensitive to the magnetic field and the shock velocity.    

For the system \replaced{analyze}{analyzed} here the size of the acceleration region \replaced{constraints}{constrains} the maximum energies, \replaced{given}{giving} $E_{\rm max} \sim 1.3 $\,TeV for electrons and  protons, with $B|_{\rm shock} \sim 0.7\,{\mu}$\,G. Note that these values are different when  \deleted{consider} other values for the magnetic field \added{are considered} (see Sect.\ref{sec:magdep}).

\added{\subsection{Time scales}}

\begin{figure}
\begin{center}
\includegraphics[width=\columnwidth,trim=.20cm .0cm .50cm .20cm, clip=true,angle=0]{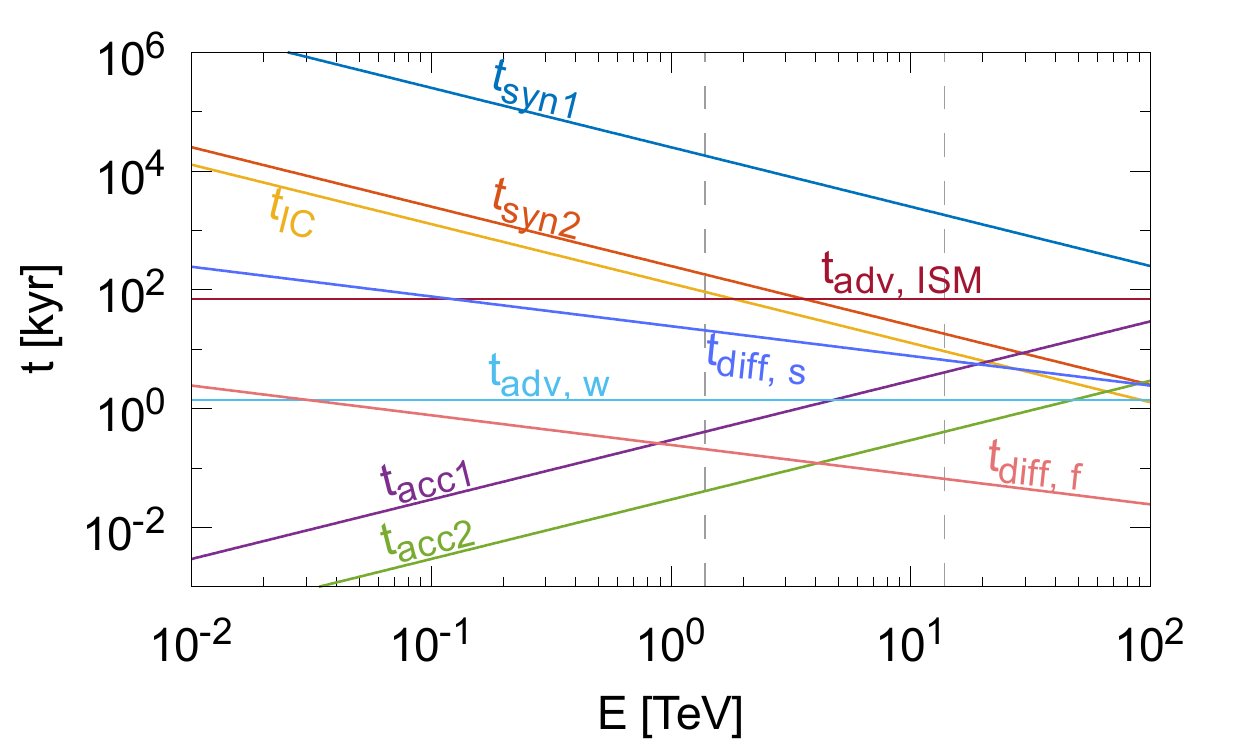}
\caption{\added{Time scales involved in the physical processes of the system as a function of the electron energy. The plotted scales are: the synchrotron cooling time for two values of the stellar magnetic field, the cooling time for the IC scattering with dust photons for $U_{\rm} = 10^{-11}$\,erg\,{cm}$^{-3}$, the acceleration time for two values of the stellar magnetic field, the diffusion time for the slow and fast cases, the  advection time for the wind and the ISM. The vertical dashed lines show the maximum energies imposed by the system size for two values of the stellar magnetic field. See the text for further details.}}
\label{fig:tscale}
\end{center}
\end{figure}

\added{The time scales discussed so far are plotted in Figure\,\ref{fig:tscale}, as a function of the electron energy. The synchrotron losses are shown for two values of the magnetic-field strength, $B|_{\rm shock} = 0.7$ and $B|_{\rm shock} = 7$\,$\mu$G, with the subindexes 1 and 2, respectively (see Sect.\,\ref{sec:magdep}). This corresponds to two  stellar magnetic field values $B_{\star} = 10^{2}$ and $10^{3}$\,G. The acceleration time is also plotted for the two $B|_{\rm shock}$ values. The IC losses
\footnote{We consider here only the Thomson regime, which is appropriate for the energies of interest.} due to dust photon scattering are plotted for a representative constant value of $U_{\rm}$ of $10^{-11}$\,erg\,{cm}$^{-3}$ (see Fig.\,\ref{fig:energydensity}).}

\added{The time scales for the transport processes are also presented in Figure\,\ref{fig:tscale}.  For the diffusion we plot the time scale for the two cases studied here: fast and slow. We show the cases of advection produced by the wind and by the ISM. The dashed vertical lines indicate the maximum energy arising from the constraint imposed by the size of the system, for the two values of $B|_{\rm shock}$ considered  (see Sect.\,\ref{sec:maxener}).}

\added{From the Figure we can conclude that transport effects are of great significance and dominate over the losses, at least at the injection position. Fast diffusion dominates the transport for energies greater than 50\,GeV. In the case of slow diffusion it dominates the transport over the ISM advection for energies greater than 0.2\,TeV and the wind advection for very high energies (i.e. $E >$ 265\,TeV).}

\subsection{Calculation details}

\deleted{The}Equation\,~(\ref{eq:transport}) is solved using a discrete grid  
$(E, r, z) \in [ 1 \, {\rm keV}, \, 10 \, {\rm TeV} ] \times [ 0, \, 24 \, {\rm pc} ] \times [ -12, \, 10 \, {\rm pc}]$, using the finite-volumes method. The energy grid is \replaced{logarithmic}{logarithmically} spaced and the spatial grid is uniform. The used  grid resolution \added{for $(E, r, z)$} is $(L,M,K) = (128,128,128)$\added{, respectively}.  Particles are injected through all the integration time. The resulting $N(t,\,E, \,r, \,z)$ for electrons and protons are interpolated into a 3D spatial grid. We calculate the non-thermal radiation produced by the particles as they diffuse through the domain. The integration proceeds until there are no significant changes in the radiation outcome.

Initially we assume $N(0,\,E, \,r, \,z) \equiv 0$, i.e. no particles inside the domain.  The energy boundary conditions are $N(t,\,E\,>\,E_{\rm max}, \,r, \,z) = 0$ and  $N(t,\,E\,<\,E_{\rm min}, \,r, \,z) = 0$\replaced{; this}{This} does not influence the system evolution, because the upper limit is above the maximum energy of the injected particles, and the advection in the energy space is always \replaced{direct}{directed} to smaller energies. The outer boundary condition for $r$ and the inner and outer boundary conditions for $z$  are assumed as outflow; also no inflow is allowed at the inner $z$ boundary. We adopted axial symmetry at the $r$ inner boundary. 

The numerical integration is performed through the operator splitting method. Each time-step integration \replaced{evolved}{computes} the \added{evolution of the} particle density distribution on the grid through four sub-steps: first the losses are integrated, then the spatial advection followed by spatial diffusion and finally the source term is added. The time-steps were chosen in accordance with the CFL stability criterion. Further description of the code is made in \citet{2015MNRAS.448..207D,2017arXiv171106250D}.

\section{Results}\label{sec:results}

\begin{figure*}[t!]
\begin{center}
\includegraphics[scale=.8, clip=true,angle=270]{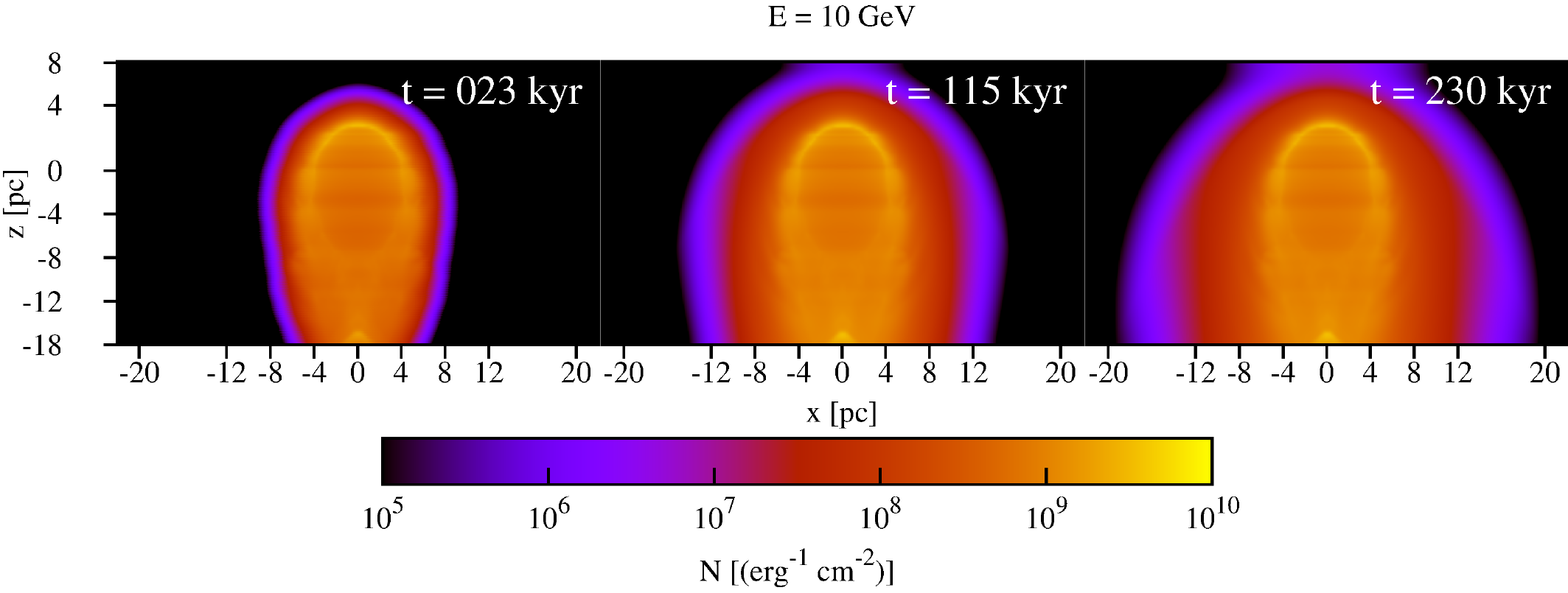}
\caption{Electron distribution at fixed energy projected along the line of sight for the slow diffusion case; $E_e = 10$\,GeV. Time evolves from left to right. }
\label{fig:bolometricemap}
\end{center}
\end{figure*}

\begin{figure}
\begin{center}
\includegraphics[scale=.3, trim=0.21cm 6.9cm 0.2cm 13.4cm, clip=true,angle=0]{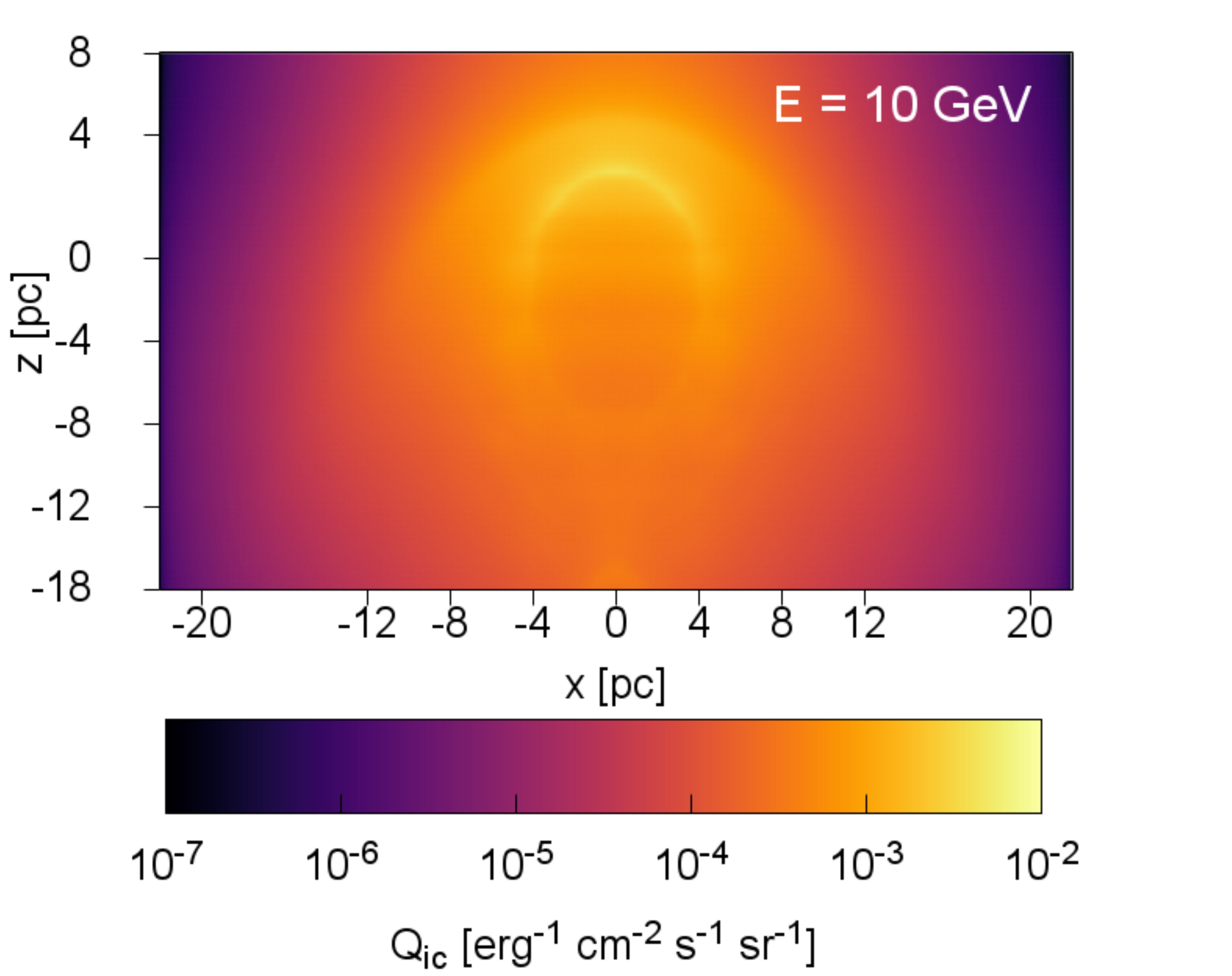}\\
\includegraphics[scale=.3, trim=0.21cm 6.9cm 0.2cm 13.4cm, clip=true,angle=0]{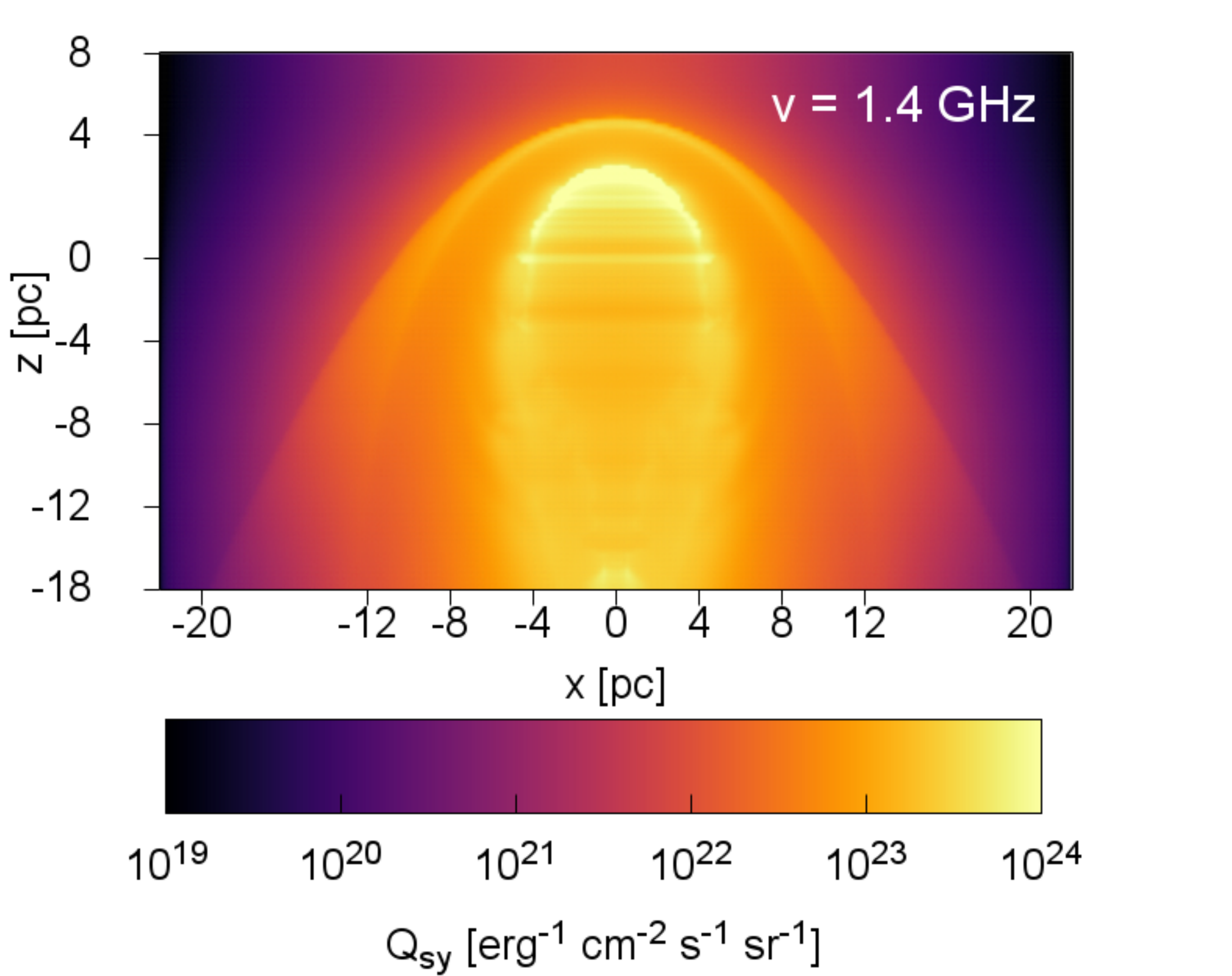}
\caption{Intensity evolution, projected along the line of sight for IC at $E = 10$\,GeV (up) and synchrotron at $E \equiv 1.4\,$GHz (bottom); the figures correspond to $t = 230$\,kyr.}
\label{fig:bolometricqmaps}
\end{center}
\end{figure}

In the Figure\,\ref{fig:bolometricemap} it is shown a map of the distribution of electrons for $E = 10$\,GeV, for different evolution times. The 2D maps are constructed integrating the 3D data along an arbitrary line of sight, chosen here to be on the $y$-direction. This plot corresponds to the slow diffusion case  $D_{\rm 10\,GeV} = 10^{25}$\,cm$^{2}$\,s$^{-1}$. The integration time or injection time $t_{\rm inj}$ is taken as 230\,kyr\replaced{, for this time a typical particle injected at $z \sim$ -9\,pc would have crossed the bottom boundary}{. This time is enough for a particle injected at $z \sim$ -9\,pc to cross the bottom boundary}. From the maps it can be seen that particles are injected in the reverse shock position, and then advected and diffused in the plane. The maximum number of electrons is always near the injection region and when  $t = 230$\,kyr all the domain is reached by particles\replaced{;}{.} \added{Only a} few particles, the most energetic ones,  reach the wind region, \added{and} most of them are advected away by the wind. There are some bright spots in which  particles are accumulated, because the velocity is very low in these regions and  diffusion is slow (see Fig.\,\ref{fig:velocitymap}). In the Figure\,\ref{fig:bolometricqmaps} upper plot we show the IC map for $E = 10$\,GeV at the final time $t = 230$\,kyr. The maximum emission occurs in the vicinity of the reverse shock\replaced{; it}{. It} becomes stronger in the region above the injection position as the electrons reach by diffusion the regions of highest   $U_{\rm ph}$ (see Fig.\,\ref{fig:energydensity}), slightly  tracing the bow shock structure. This last effect is stronger for the  synchrotron\replaced{,  which}{emission whose}  map at $E \equiv 1.4$\,GHz is shown in the bottom plot of Figure\,\ref{fig:bolometricqmaps}\replaced{; the}{. The} behavior exhibited by this emission is similar: the maximum here occurs in the shocked wind region and then in the shocked ISM.  

\begin{figure}
\begin{center}
\includegraphics[scale=.37,trim=1.cm 0.cm 1cm 1cm, clip=true,angle=270]{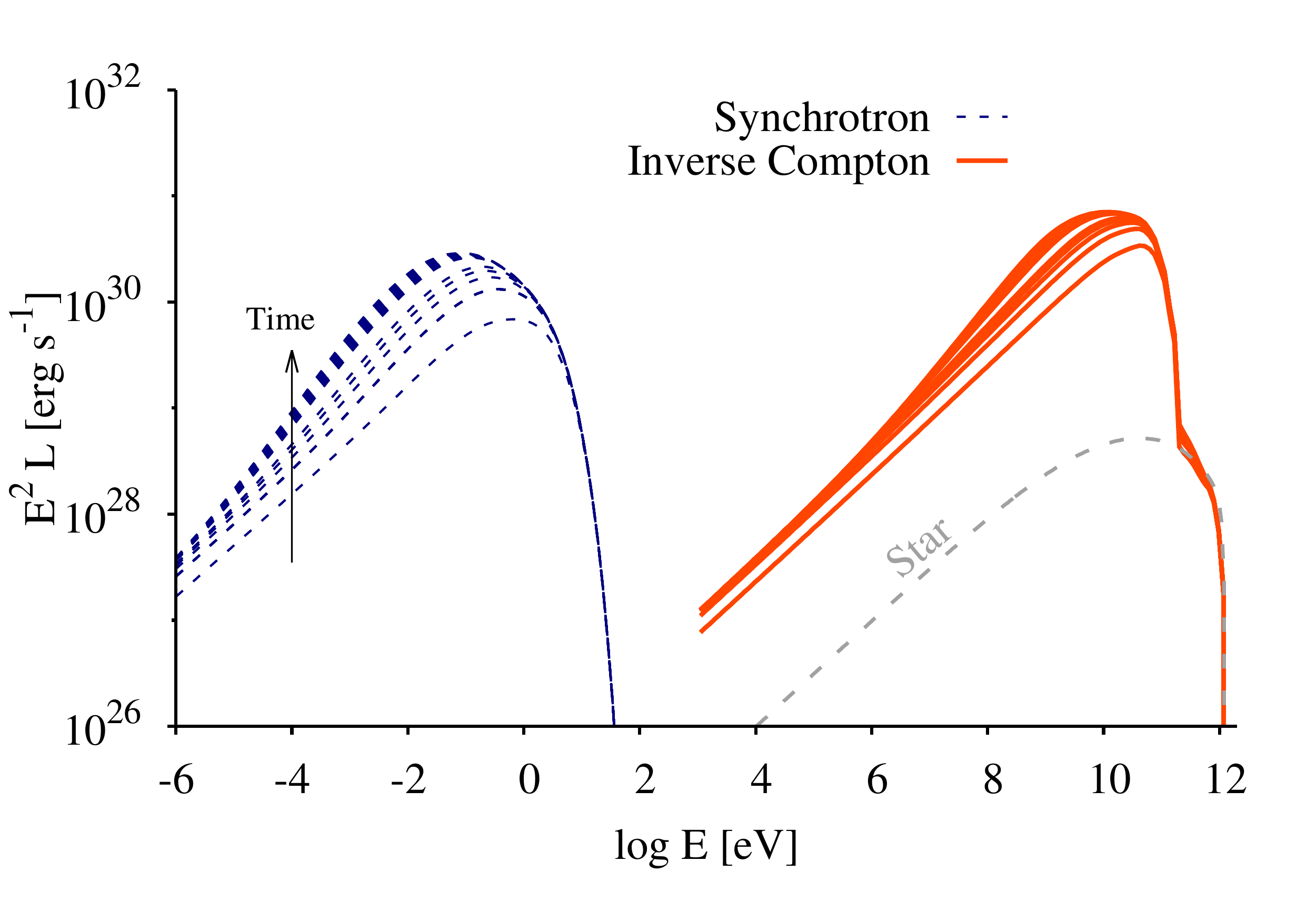}
\caption{Spectral energy distribution for different integration times from the bow shock of a massive runaway star. The grey curve illustrates the contribution to the IC of the stellar photons. The black arrow indicates that the radiation grows with time.}
\label{fig:sed1}
\end{center}
\end{figure}

The volume integration of the non-thermal luminosity --the SED-- for the slow diffusion case is shown in Figure\,\ref{fig:sed1}. Only the dominant processes are shown:  IC scattering and synchrotron. The luminosity grows with time, as indicated with the black arrow in the plot. However after some time the emission stop growing\replaced{, this}{. This} is because a steady state is reached between the injection, advection, losses and diffusion of particles in the domain. This can be appreciated in the pileup of the curves in the SED as time passes.  The two \replaced{components of the IC}{IC components}, from the star and from the dust emission, can be distinguished in the curves\replaced{; for}{. For} illustration we have plotted the contribution  from the stellar photons in grey. \replaced{This}{As can be seen in Fig.\,\ref{fig:sed1} this} component is rather weak.

The emission from interactions with matter (relativistic Bremsstrahlung and $p-p$ inelastic collisions)  is very low when compared with IC, with maximum luminosities $\sim$ $10^{30}$\,erg\,s$^{-1}$; hence the hadronic contribution to the emission is unimportant and the relativistic protons diffuse out of the system almost without energy loss as predicted previously \citep{2015MNRAS.448..207D}. We are not discussing these emission components any further.

The spectrum of the resulting SED depends, among other factors,  on the shape of the injected particles. \replaced{ and a change on the spectral index changes the distribution of photons at specific energies}{A change on the spectral index will alter the photon distribution}. In a DSA process at a non-relativistic shock  \added{we expect} $\alpha \sim 2$, but the spectral index can \replaced{be deviated}{deviate} from that value \citep[see e.g.,][]{2011hea..book.....L}. In the case of an injection $\alpha = 1.8$ \added{the emission through all the spectrum diminishes a factor $\sim$ 2.2 due to a change in the particles normalization ($a \sim 90$, see Sect.\,\ref{sec:injection}). Also, the distribution of radiation in the SED changes and} more emission is produced at the highest energies; on the contrary the radiation diminishes at radio and X-rays. At the highest energies the shape of the SED is modified by diffusion, and its shape \replaced{depends on}{is influenced by} the dependence of the diffusion coefficient with energy. 

The gamma emission coming from $z>0$\replaced{, this is the apsis of the bow shock,}{(the apsis of the bowshock)} dominates the radiation output. For example, for  $E = 100$\,GeV  the IC  \replaced{luminosity per unit parsec $L_{\rm IC}(100 {\rm GeV})\,|_{z>0}\,/8\,$pc}{intensity at $z = 8$\,pc is twice that at $z = 26$\,pc} \deleted{doubles the total emission per unit parsec $L_{\rm IC}(100 {\rm GeV})\,|_{\forall z}\,/26\,$pc}, meaning that the radiation density is higher in this region, and the bulk of the emission is coming from here. This is because the IR target radiation field is strong and the injection is higher in this region.

\subsection{Dependence on diffusion}

\begin{figure}
\begin{center}
\includegraphics[scale=.3,trim=2cm 1.5cm 2.1cm 1.cm, clip=true,angle=270]{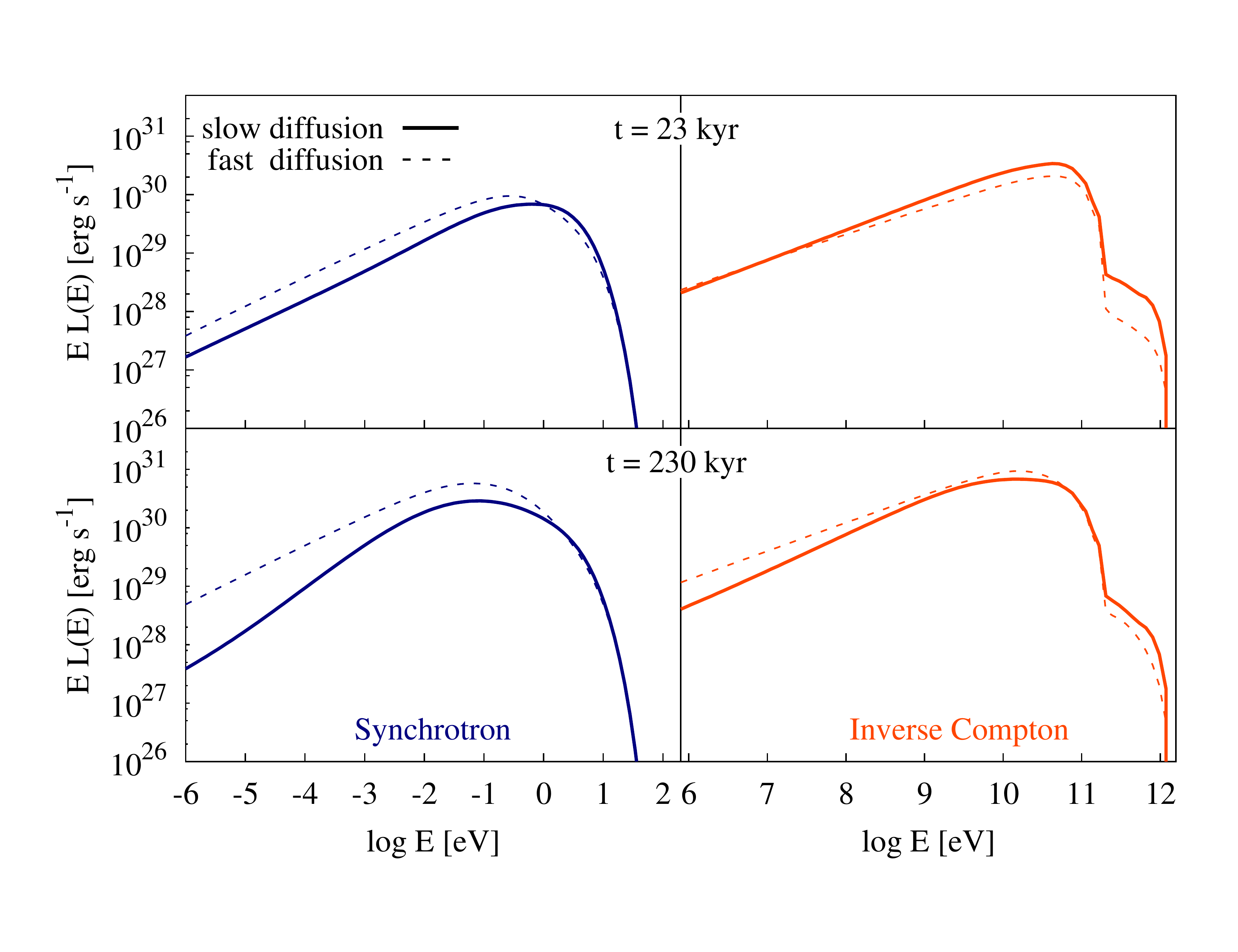}
\caption{Spectral energy distribution, at different integration times,  for different diffusion regimes: slow ($D_{10\,{\rm GeV}} = 10^{25}$\,cm$^{2}$\,s$^{-1}$) and fast ($D_{10\,{\rm GeV}} = 10^{27}$\,cm$^{2}$\,s$^{-1}$).}
\label{fig:sed3}
\end{center}
\end{figure}

We consider slow ($D_{10\,{\rm GeV}} = 10^{25}$\,cm$^{2}$\,s$^{-1}$) and fast ($D_{10\,{\rm GeV}} = 10^{27}$\,cm$^{2}$\,s$^{-1}$) diffusion to study how these different regimes affect the non-thermal luminosity. In the plots of Figure\,\ref{fig:sed3} we show the dominant non-thermal components at $t = 23$\,kyr (up) and at $t = 230$\,kyr (bottom). Initially the differences between the cases are not so important, but in the IC star component, which is stronger in the case of slow diffusion, the particles stay \replaced{more}{for a longer} time in the vicinity of the injection region where the stellar radiation field is stronger.  The fast diffusion slightly dominates the synchrotron component for $E < 1\,$eV; this is because the high-energy particles reach regions of stronger magnetic field (see Fig.\,\ref{fig:magneticfield}). At the final integration  time the fast  diffusion  dominates the SED\added{, except for the IC star component}; this is because particles in the slow diffusion are dragged by advection and very few reach the regions of highest magnetic and \replaced{photon fields}{IR photon field}; as discussed in Sect.\,\ref{sec:advection}, in the slow diffusion case due to advection a typical particle will not reach the denser bow shock.

\subsection{Dependence on magnetic field}\label{sec:magdep}

Here we assume a greater value for the stellar magnetic field, with $B_{\star} = 1$\,kG. This might affect \added{the magnetic field in the wind region  and it might change the hydrodynamics of the system changing, among other things, the reverse shock position. If the magnetic pressure is of the same order or higher than the wind ram pressure, the pressure balance (see Sect.\,\ref{sec:intromodel}) between the wind and the ISM occurs at greater distances from the star.} 

\added{The relevance of the effects of the magnetic field on the hydrodynamics can be estimated by comparing the ram pressure $\rho_{\rm w}V_{\rm w}^{2}$ with the magnetic pressure of the wind $B_{\rm wind}^2/8\pi$. The wind ram pressure is given by:
\begin{equation}
P_{\rm r\,,w} = \frac{\dot{M_{\rm w}}}{4 \pi R^{2} V_{\rm w}}V_{\rm w}^{2} \equiv \frac{p_{\rm r\,w}}{R^{2}};   
\end{equation}
the last factor of  Eq.(\ref{eq:magwind}) tends to 1 very fast for $R > 20\,R_{\star}$, then the magnetic pressure is:
\begin{equation}
P_{\rm B\,,w} = \frac{B_{\star}^2}{8 \pi}\left[ 1 + \left(\frac{V_{\rm w}}{V_{\rm rot}}\right)^{2}\right]^{-1}\left(\frac{R_{\rm star}}{R}\right)^{2} \equiv \frac{p_{\rm B\,w}}{R^{2}}.
\end{equation}
\noindent Then the pressures ratio   $P_{\rm r\,,w} / P_{\rm B\,,w} \sim 7$ for this case, and the magnetic field in the wind is not expected to affect dramatically the hydrodynamics, and can be ignored at least at first order.  }

The magnetic field at the reverse shock $B|_{\rm shock}$ \replaced{, that growths}{grows by} one order of magnitude\replaced{;}{,} hence $t_{\rm acc}$ \added{increases} and \deleted{the maximum energy that} electrons reach \added{a higher maximum energy} in the acceleration process\replaced{, see Sect.\,\ref{sec:particles}}{ (see Sect.\,\ref{sec:particles})}. In this case we obtain  $E^{e}_{\rm max}$,\deleted{$\sim 9.2\,$TeV and  } $E^{p}_{\rm max}$ $\sim 13.8\,$TeV.

No great differences in the IC radiation occur, but the ones expected from the change in $E^{e}_{\rm max}$, that  is \replaced{ more than 3}{10} times higher than in the reference case. The gamma spectrum is shifted towards higher energies, increasing the total emission output (see Sect.\,\ref{sec:discussion}). Naturally the synchrotron emission is much higher; it dominates the  SED in \deleted{the} X-rays \deleted{region} until $E$ $\sim$ \replaced{10}{6.3}\,keV; with $L_{\rm syn} \sim 8\times10^{29}$\,erg\,s$^{-1}$ at $E \sim 1$\,keV and $L_{\rm syn} \sim 2\times10^{27}$\,erg\,s$^{-1}$ at $E \sim$ \replaced{10}{6.3}\,keV.

\subsection{Dependence on the stellar velocity}

The bow shock size and shape of the same type of star changes with $v_{\star}$ \citep[see e.g.,][]{2014MNRAS.444.2754M}. In \replaced{other}{order} to study the impact of this in the non-thermal emission we consider here the same massive star described in Sect.\,\ref{sec:hydro}, but with a higher velocity:  $v_{\star} = 70$\,km\,s$^{-1}$. For this system the global steady state is reached at $\sim$ 4\,Myr. For numerical stability here we use an Harten-Lax-van Leer  solver. As \deleted{we} can be deduced from the expression of $R_0$, see Eq.\,\ref{eq:R0}, the whole bow shock structure is smaller.

The injection region is closer to the star, hence the magnetic field near the injection region has greater values. The maximum energy particles might achieve is slightly higher than the previous case, with $E^{e}_{\rm max} \sim 2$\,TeV. We compute the SED for slow diffusion. The  synchrotron emission reaches higher energies than in the reference case, as a combination of a greater magnetic field near the injection region and a slightly higher electron maximum energy. Both the synchrotron and IC emission are higher in this case, by  a maximum factor of 4 at same energies. This is because the maximum values of the target fields are closer to the injection region, hence particles lose energy more efficiently (see further discussion in Sect.\,\ref{sec:discussion}).

\subsection{Synchrotron emission from the tail} 
The bow shock tail can extend for several parsecs towards the $-z$ direction. The escaping electrons would produce further synchrotron emission when interacting with the magnetic field of the shocked material in the bow shock tail. In order to evaluate how important is the emission produced further down stream we compute the emission coming from the bottom region, $-18<z<-17$\,pc, as an upper limit (further down the number of particles would be more diluted due to diffusion, and the emission per pc would be lower). 

The luminosity in this bottom region is a fraction $3\times10^{-2}$  of the total one for both fast and slow diffusion; in particular at $1.4$\,GHz, the frequency of large-area radio surveys such as FIRST \citep{1995ApJ...450..559B} and NVSS \citep{1998AJ....115.1693C}, the luminosity is some factor of $10^{27}$\,erg\,s$^{-1}$ for the first case, and $10^{26}$\,erg\,s$^{-1}$ for the other. For the fast case this value \added{is of the order of} \deleted{lies over} the radio detection limits of $1-2.5$\,mJy that is, for a source located at 1\,kpc, $1-2.5\times10^{27}$\,erg\,s$^{-1}$ \deleted{, and for the slow case is of the same order}. For the case of fast diffusion the synchrotron emission from the tail might be important and  detectable for sources at these distances or less; however it would not be higher than the emission coming from the bow shock region. A proper calculation of this contribution is beyond the scopes of this work and would be \replaced{study}{studied} elsewhere.

\section{Discussion}\label{sec:discussion}

\added{The gamma-ray photons produced in the system can be absorbed by  lower energy photons through photon-photon annihilation. The low-energy photon field can be in the source itself or in the propagation path of the gamma ray on its way to the observer. For a Galactic source this last component is negligible, and we focus on photon fields within the bow shock. For the process to occur the energies of the involved photons must fulfil\footnote{We assume here head-on collisions.}:
\begin{equation}
E_{\gamma}\epsilon > 2(m_e c^2)^2,
\end{equation}
where $E_{\gamma}$ is the energy of the gamma ray and $\epsilon$ the energy of the target photon. For $E_{\gamma} = 100$\,GeV, using $\epsilon \sim 3/2k_{\rm B}T$, $T > 40475$\,K. This means that the stellar photon field can absorb gamma rays above 100\,GeV. This is not surprising, a massive star photon field is known to be a significant source of gamma-ray annihilation, for example in a high-mass microquasar \citep[e.g.,][]{2010A&A...518A..12R} or colliding-wind binaries  \citep[e.g.,][]{2011A&A...530A..49B}. The total absorption depends strongly on the geometry and on the relative positions of the gamma ray, the stellar photon field and the observer. However we can make order of magnitude estimates.  The optical depth for a gamma ray traversing a distance $d$ is $\tau_{\gamma-\gamma} \sim \sigma_{\gamma-\gamma}n_{\star}d$. Here $n_{\star}$ is the number of photons per unit volume, that decreases quadratically with distance. The gamma-gamma cross-section maximum is $\sim \sigma_{\rm T}/5$, with $\sigma_{\rm T}$ the Thomson cross-section, and it occurs close to the threshold energy. We can estimate the maximum $d$ such that $\tau_{\gamma-\gamma} \gtrsim 1$:
\begin{equation}
\tau_{\gamma-\gamma} \sim \frac{\sigma_{\rm T}}{5}\frac{\sigma T_{\star}^4 R_{\star}^{2}}{\epsilon \,c\,d^{2}}d\,\gtrsim 1.
\end{equation}
\noindent The last condition gives $d \lesssim 10^{-4}$ light-years ($\equiv 68\,R_{\star}$). Comparing this distance with the typical scale of the system we get $d/R_{0} \sim 10^{-5}$. Even in the least favourable case the region in which the gamma-gamma absorption is important is extremely small compared to the size of the bow shock. Then, given the large extension of the gamma-ray source the absorption produced by the stellar photon field is negligible.
}

\graphicspath{./}
\begin{figure}
\begin{center}
\includegraphics[scale=.7,trim=0.21cm 1.7cm 0.1cm 0.1cm, clip=true,angle=0]{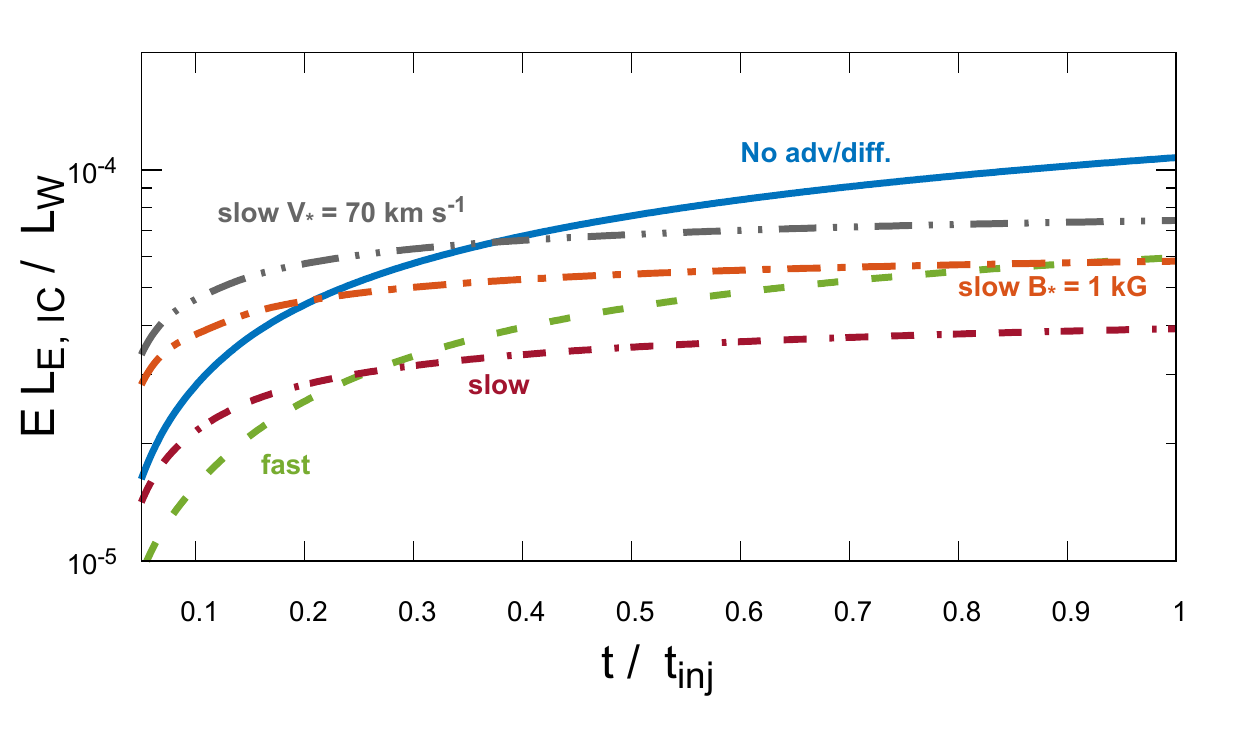}
\includegraphics[scale=.7,trim=0.21cm 1.7cm 0.1cm 0.52cm, clip=true,angle=0]{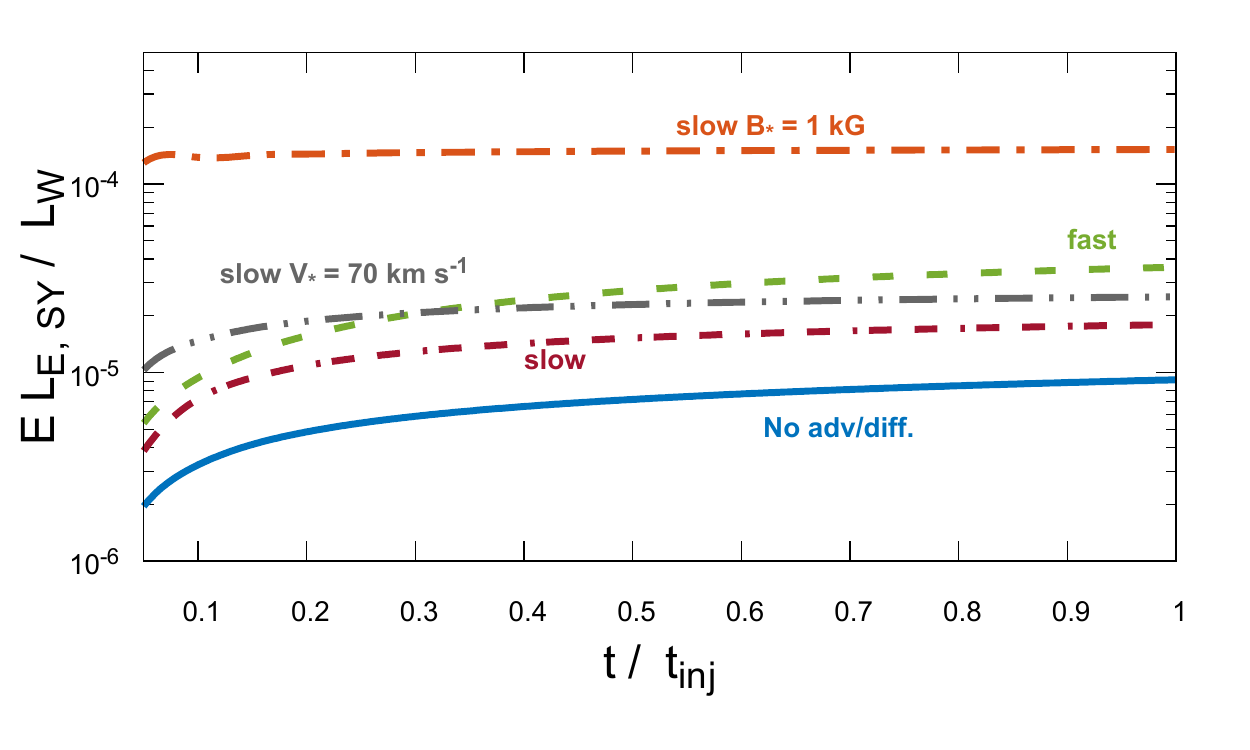}
\includegraphics[scale=.7,trim=0.21cm 0.5cm 0.1cm 0.52cm, clip=true,angle=0]{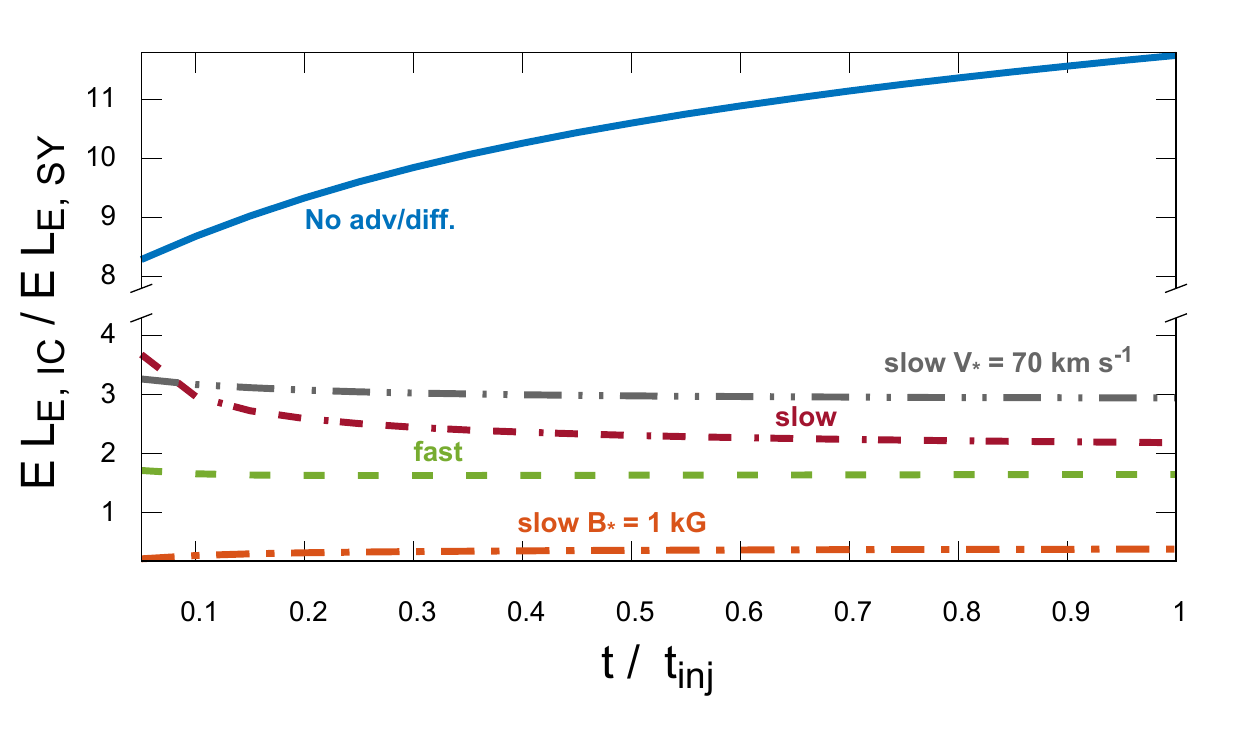}

\caption{Evolution of the ratio of the total power produced by IC scattering (up) and synchrotron radiation (\replaced{bottom}{middle}) to the wind's power $L_{\rm w}$.\added{The bottom plot shows the evolution of the IC-synchrotron power ratio.}}
\label{fig:power}
\end{center}
\end{figure}

The interaction of the stellar wind with the ISM deposits a fraction of the wind power into the ambient medium in different forms of energy. It is interesting to see then how much of the wind's power is converted into non-thermal emission as high-energy emission through IC scattering and low-energy radiation via synchrotron. This fraction as a function of the computational time is presented in Figure\,\ref{fig:power} (this depends on the assumptions made in $L_{\rm rel}$, see Sect.\,\ref{sec:injection}). In this figure it is shown the ratio of the IC emission between $1$\,keV and  $E^{e}_{\rm max}$ (upper plot) to the wind's power $\chi_{\rm IC}$,  for  the models we consider in this work; we also show a case without diffusion or advection  just for comparison. The most efficient case is that with no transport of particles, with $\chi_{\rm IC} \sim$ \replaced{2$\times 10^{-3}$}{10$^{-4}$} (blue line) at the final time, this is expected because particles stay in the box only losing energy by radiative losses. Follows by the case of the star with a higher \replaced{spacial}{spatial} velocity with a power ratio of \replaced{1.4$\times 10^{-3}$}{7.4$\times 10^{-5}$} (grey line). Next comes the case with $B_{\star} = 1$\,kG   (orange line) with \replaced{$1.2\times10^{-3}$}{5.8$\times 10^{-5}$}, the case of fast diffusion (green line) \replaced{leis}{lies}  slightly below\replaced{; finally}{. Finally} the slow diffusion case (red line) that gives $\chi_{\rm IC} \sim$ \replaced{7$\times10^{-4}$}{{4$\times 10^{-5}$}}, only 3 times less than the best case. We can see that propagation effects are important. Initially all the cases show differences in $\chi_{\rm IC}$, but as time evolves  all cases reach values $\sim$ \replaced{$10^{-3}$}{{5$\times 10^{-5}$}}.

In the \replaced{bottom}{middle} plot of Figure\,\ref{fig:power}  it is shown the ratio of the synchrotron radiation integrated between $10^{-6}$\,eV and  $10^{5}$\,eV to the wind's power: $\chi_{\rm S}$. The most efficient cases do not coincide with those of the IC discussed above\replaced{; the}{. The} synchrotron emission is very sensitive to the magnetic field value and in this highly spatially changing environment the propagations effects are really important. This can be \replaced{notice}{noticed} when analyzing the case of no diffusion and no advection, the efficiency is the lowest with a ratio to the wind's power of \replaced{2$\times10^{-4}$}{{9$\times 10^{-6}$}}. Small differences are exhibited at the final times in the case of slow and fast diffusion, with  \replaced{3$\times10^{-4}$}{2$\times10^{-5}$} for the slow and \replaced{7$\times10^{-4}$}{4$\times10^{-5}$} for the fast case: once the particles reach by diffusion the regions of higher magnetic field they radiate more effectively. When the particles are injected in a region where the magnetic field is higher the particles radiate more power, as in the case of a smaller bow shock (\replaced{gray}{grey} curve), which $\chi_{\rm S}$ is \replaced{5$\times10^{-4}$}{2.5$\times10^{-5}$}. \deleted{Obviously the} \added{The extreme} case with a higher stellar magnetic field is very efficient, with $\chi_{\rm S}\sim$ \replaced{3$\times10^{-3}$}{$10^{-4}$} \deleted{this is an extreme case}. Again for the final time there are no big differences in the energy injected as synchrotron radiation, that is  $\chi_{\rm S}\sim$ \replaced{5$\times10^{-4}$}{3$\times10^{-5}$}.

\added{From the value of $\chi_{\rm IC}$ we can also estimate what fraction of the injected power is radiated in the IC process: $L_{\rm IC}/L_{\rm rel} \sim \chi_{\rm IC}/\xi$ $\sim$ $8\times 10^{-4} - 2\times 10^{-3}$. For the synchrotron  we get a slightly smaller number (see next paragraph). We can see that electron radiation is not efficient and the electrons are transported out of the system by diffusion and advection  before they can lose a significant fraction of their power, in accordance with the time-scale estimations presented in Figure\,\ref{fig:tscale}.}

\added{We analyze also the evolution of the power ratio between the two dominant radiation mechanisms. This ratio is shown in the bottom plot of Figure\,\ref{fig:power} for all the configurations studied here. The synchrotron power only dominates in the case of a higher stellar magnetic field by almost one order of magnitude. In general the ratio between the emitted powers is of order unity, dominated by IC, with the exception of the case with no transport effects. In this case the IC power is one order of magnitude greater than the synchrotron power. Again  consistent with the time scales shown in Fig.\,\ref{fig:tscale}.} 

From the above analysis we can infer very generally the following:

\begin{eqnarray}\label{eq:lumi}
L_{\rm IC} & \lesssim & 10^{32} \left(\frac{\dot{M}}{10^{-6}M_{\odot}{\rm yr}^{-1}}\right)\left(\frac{V_{\rm w}}{2000\,{\rm km}\,{\rm s}^{-1}}\right)^{2}\times \nonumber \\
&& \left(\frac{\chi_{\rm IC}}{10^{-4}}\right)\left(\frac{40}{a}\right)\left(\frac{\xi}{0.05}\right) {\rm erg}\,{\rm s}^{-1}, \nonumber\\
L_{\rm Sy} & \lesssim & 5 \times 10^{31} \left(\frac{\dot{M}}{10^{-6}M_{\odot}{\rm yr}^{-1}}\right)\left(\frac{V_{\rm w}}{2000\,{\rm km}\,{\rm s}^{-1}}\right)^{2} \times \nonumber \\
&& \left(\frac{\chi_{\rm S}}{5 \times 10^{-5}}\right)\left(\frac{40}{a}\right)\left(\frac{\xi}{0.05}\right){\rm erg}\,{\rm s}^{-1}.
\end{eqnarray}

\noindent \added{We explicitly show the dependence with the shock efficiency $\xi$ and proton-to-electron power ratio $a$.}In the case of IC the maximum power is around $\sim 100$\,GeV with a luminosity of approximately 10\% of the above value. These are  modest values for a gamma-ray source. For the synchrotron, the maximum luminosity lies around $\sim 1$\,eV, ignoring the extreme case with $B_{\star} = 1$\,kG.\added{The value of $a$  changes in two cases. If the injection index changes (this was discussed previously): in the case of a softer index ($\alpha > 2$)  $a$ decreases and it increases for harder indexes ($\alpha < 2$). The other case is if the condition for equal injected number rate for both species is relaxed, smaller values can occur if more electrons are injected. The shock efficiency $\xi$ adopted here is modest, it can be higher, between $10$ to $20$\%, as obtained in numerical simulations \citep{2014ApJ...783...91C} or observations of the earth's bow shock \citep{1990ApJ...352..376E}.}

\added{We can apply Eq.(\ref{eq:lumi}) to the case of Lambda Cep associated with a {\it Fermi} source \citep{2018arXiv180600614S}.  For $\dot{M} \sim 7\times 10^{-6}\,M_{\odot}\,$ yr$^{-1}$ and $V_{\rm w} \sim 2200$\,km\,s$^{-1}$ \citep{2007A&A...473..603M}, we get $L_{\rm IC}$  $\lesssim  8.5\times10^{32}\left(\frac{40}{a}\right)\left(\frac{\xi}{0.05}\right){\rm erg}\,{\rm s}^{-1}$. This result is consistent with the gamma source power at 100\,GeV of $\sim$ $10^{32}$\,erg\,s$^{-1}$. The case of LS 2355 is more complex because the system is interacting with a HII region.}

According to our model the maximum emission from massive runaway bow shocks is not to occur in the very high energy domain, i.e. $>$ TeV.  Our results from Figure\,\ref{fig:power}  are in agreement with H.E.S.S. upper limits, i.e. $L_{\rm IC} [0.14-18\,{\rm TeV}]$ $<$ $10^{-2}$\,$L_{\rm W}$ \citep{2017arXiv170502263H}. Concerning the upper limits from {\it Fermi} from \citet{2014A&A...565A..95S}, although they are for specific sources, for those investigated in \citet{2017MNRAS.471.4452D} with distances ranging between 200 and 2000\,pc, these upper limits range between $10^{33}$ to $10^{35}\, $erg\,s$^{-1}$ in the 4 energy bands. Not even our most favorable model at gamma rays   reach these upper limits; however for the case of a more powerful wind it might reach these levels (see Eq.\ref{eq:lumi}). In general, the distances of the bow shocks cataloged in the E-BOSS \citep{2012A&A...538A.108P} also ranged between $\sim$ 200 and 2000\,pc;  the theoretical 5$-\sigma$ sensitivity of {\it Fermi} in the energy range between  1 and 10\,GeV  is $\sim$ $10^{-11}$\,erg\,s$^{-1}$\,cm$^{-2}$ (it can be smaller for sources above the plane). For sources at  200 and 2000\,pc the threshold luminosity is $\sim$ $5\times10^{31}$ and  $5\times10^{33}$\,erg\,s$^{-1}$, respectively. These values are not unrealistic for our model\replaced{; however}{. However}  a note of caution is in order: the power in relativistic electrons might be supper estimated, as can be learned from the radio upper limits as discussed below. 

The $3-\sigma$ radio upper limits in the case of the sources from the study of \citet{2017MNRAS.471.4452D} are more restrictive than those at gamma rays. These upper limits are obtained from the NRAO VLA Sky Survey (NVSS), a 1.4\,GHz ($\sim 5.8\times 10^{-6}$\,eV) continuum survey. These values range between $10^{27}$ to $10^{28}$\,erg\,s$^{-1}$. If these limits are \replaced{to apply}{applied} to a system like the one we are studying here, then the synchrotron power we obtain with our models is \added{roughly} over these limits \deleted{for a factor 10 , or more in case of $B_{\star} = 1$\,kG}. This means that in the presence of a relatively high magnetic field the power in electrons assumed here could be overestimated by at least the same factor. If this is the case, then the  IC luminosity is lower than the one predicted in our models. Another possibility is that the magnetic field is over estimated.

Very low values of \added{the} magnetic field are not good either for producing higher values of gamma emission\replaced{, particles}{. Particles} need magnetic field to be efficiently accelerated in the reverse shock to high energies. A weak magnetic field  would not produce electrons energetic enough to produce gamma rays (see Sect.\,\ref{sec:injection}). \added{An electron to emit synchrotron radiation at a frequency $\nu$ needs an energy $E = 7.9 \left(\nu/[{\rm GHz}]\right)^{1/2} \left(B/[\mu{\rm G}]\right)^{-1/2}\,{\rm GeV}$ \citep[e.g.,][]{1970ranp.book.....P}, for $\nu = $ 1.4\,GHz in the ISM magnetic field $E \sim 4$\,GeV. Hence a strong radio signal at these frequencies does not necessarily imply the presence of relativistic electrons capable of producing gamma radiation at energies higher than 10\, GeV.}

Another possibility that might decrease the synchrotron at 1.4\,GHz without assuming a smaller power in relativistic particles is that the injected electrons have a \replaced{stepper}{steeper} power-law index, i.e. $ |{\alpha}| < 2$, as we learn from the previous Section. With a steep injection the emission at long wavelength decreases. A smaller value of $|{\alpha}|$ then would give a steeper photon distribution, decreasing the emission in the energy region of interest. However this effect is not expected to \replaced{produced}{produce} dramatic changes. \added{It is worth mentioning that given the sensitivity of present observatories the lack of detection does not constitute a strong evidence for a lack of efficient particle acceleration in these sources.
}

The above analysis is made  extrapolating the upper limits from a sample of 5 sources to all sources, and this might not be the general case. In particular it does not apply to the case of BD +43$^{\circ}$3654, that was in fact detected at radio. In this system the emission  detected at 1.42 and 4.86 GHz is of the order of $\sim$ $10^{30}$\,erg\,s$^{-1}$. \added{Such a high luminosity is not even achieved for a higher value of the magnetic field strength (e.g. our $B_\star = 1$\,kG model)}
\deleted{We only achieve these values in the case with $B_{\star} = 1$\,kG}.

The upper limits at X-rays between 0.3 and 10\,keV from previous works are between $10^{30}$ and  $10^{31}$\,erg\,s$^{-1}$; in the cases studied here, except the case with $B_{\star} = 1$\,kG, the luminosity between 1 and 10\,keV lies below\deleted{, or is of the order of,} these values. For these cases the analysis made in \citet{2017MNRAS.471.4452D} still holds, this is: current detectors are not able to differentiate between the non-thermal emission, if any, and the stellar thermal one. The case of a high stellar magnetic field the emission should be detectable at X-rays \added{with present observatories}\replaced{, a}{. A} lack of detection might indicate that \added{such a large value for  $B_{\star}$ is not reached in these objects or that the magnetic field in the wind is overestimated.}\deleted{this value for $B_{\star}$ is unrealistic}

\section{Concluding remarks}\label{sec:conclussions}

In this work we study a very general case of a massive runaway star bow shock, assuming typical values for describing the system and ordinary assumptions. The strongest assumption that is made in our modeling is the acceleration of electrons through DSA in the wind shock. The only indirect evidence that supports this assumption is the observation of synchrotron emission from the bow shock of BD +43$^{\circ}$3654. This hypothesis \replaced{would}{will} be carefully analyzed in a future work.

In what follows we summarize the main conclusions of this study:
   \begin{itemize}   
      \item According to our model the non-thermal emission produced in the bow shock of a massive runaway star \replaced{are}{is} mainly  \added{made of} synchrotron radiation and IC emission at gamma rays, as predicted \replaced{in}{by} previous works.       
     \item In the general case the luminosity predicted here at X-rays \replaced{leis}{lies} below the existing X-ray upper limits. In the  case of a strong stellar magnetic field the synchrotron radiation is the dominant process at soft X-rays.
      \item A fraction between \replaced{$10^{-4}$ and $10^{-3}$}{$4\times10^{-5}$ and $10^{-4}$} of the wind power is converted into IC radiation; with a maximum around $E = 100$\,GeV.
       \item A fraction between \replaced{$10^{-5}$ and $10^{-3}$}{$9\times10^{-6}$ and $10^{-4}$} of the wind power is converted into synchrotron emission; with a maximum around $E = 1$\,eV. \deleted{In the extreme case of a strong stellar magnetic field this fraction can reach $10^{-3}$.}       
      \item Transport effects, advection and diffusion, \deleted{are very important for the} \added{dominate over} radiation losses \deleted{outcome}. \added{Only $\sim 0.16-0.4$\% of the injected power in electrons is radiated, the bulk of the particles leaves the system and radiates elsewhere.}      
      \item Synchrotron emission from the bow shock tail, produced by dragged electrons, might be important\added{, especially in the fast diffusion regime}.      
      \item The bulk IC radiation is coming from the \emph{cup} region of the bow shock.
      \item The  hadronic component in the SED is completely negligible; protons diffuse and advect into the ISM almost without loosing energy.
     \item Given the better sensibility of current instruments at radio wavelengths theses systems are more prone to be detected at radio through the synchrotron emission they produce rather than at gamma energies.
      \item The lack of detection at radio of specific sources put \replaced{great}{stringent} constraints in the emission expected at gamma rays.\deleted{low upper limits at these frequencies can be indicators that the bow shocks from runaway massive stars are not efficiently accelerating electrons.}
   \end{itemize}

\acknowledgments
M.~V.~d.V. acknowledges support from the Alexander von Humboldt Foundation. The authors would like to thank Dr. Reinaldo Santos-Lima for fruitful discussions. We also thank the anonymous referee for insightful comments.
\software{PLUTO \citep{2007ApJS..170..228M}}

\bibliographystyle{yahapj}

\bibliography{Yourfile}

\begin{thebibliography}{}
\providecommand\natexlab[1]{#1}
\providecommand\JournalTitle[1]{#1}

\bibitem[{{Abdo} {et~al.}(2013){Abdo}, {Ajello}, {Allafort}, {Baldini},
  {Ballet}, {Barbiellini}, {Baring}, {Bastieri}, {Belfiore}, {Bellazzini}, \&
  et~al.}]{2013ApJS..208...17A}
{Abdo}, A.~A., {Ajello}, M., {Allafort}, A., {et~al.} 2013,
  \href{http://dx.doi.org/10.1088/0067-0049/208/2/17}{\JournalTitle{\apjs},
  208, 17}

\bibitem[{{Acero} {et~al.}(2015){Acero}, {Ackermann}, {Ajello}, {Albert},
  {Atwood}, {Axelsson}, {Baldini}, {Ballet}, {Barbiellini}, {Bastieri},
  {Belfiore}, {Bellazzini}, {Bissaldi}, {Blandford}, {Bloom}, {Bogart},
  {Bonino}, {Bottacini}, {Bregeon}, {Britto}, {Bruel}, {Buehler}, {Burnett},
  {Buson}, {Caliandro}, {Cameron}, {Caputo}, {Caragiulo}, {Caraveo},
  {Casandjian}, {Cavazzuti}, {Charles}, {Chaves}, {Chekhtman}, {Cheung},
  {Chiang}, {Chiaro}, {Ciprini}, {Claus}, {Cohen-Tanugi}, {Cominsky}, {Conrad},
  {Cutini}, {D'Ammando}, {de Angelis}, {DeKlotz}, {de Palma}, {Desiante},
  {Digel}, {Di Venere}, {Drell}, {Dubois}, {Dumora}, {Favuzzi}, {Fegan},
  {Ferrara}, {Finke}, {Franckowiak}, {Fukazawa}, {Funk}, {Fusco}, {Gargano},
  {Gasparrini}, {Giebels}, {Giglietto}, {Giommi}, {Giordano}, {Giroletti},
  {Glanzman}, {Godfrey}, {Grenier}, {Grondin}, {Grove}, {Guillemot}, {Guiriec},
  {Hadasch}, {Harding}, {Hays}, {Hewitt}, {Hill}, {Horan}, {Iafrate}, {Jogler},
  {J{\'o}hannesson}, {Johnson}, {Johnson}, {Johnson}, {Johnson}, {Kamae},
  {Kataoka}, {Katsuta}, {Kuss}, {La Mura}, {Landriu}, {Larsson}, {Latronico},
  {Lemoine-Goumard}, {Li}, {Li}, {Longo}, {Loparco}, {Lott}, {Lovellette},
  {Lubrano}, {Madejski}, {Massaro}, {Mayer}, {Mazziotta}, {McEnery},
  {Michelson}, {Mirabal}, {Mizuno}, {Moiseev}, {Mongelli}, {Monzani},
  {Morselli}, {Moskalenko}, {Murgia}, {Nuss}, {Ohno}, {Ohsugi}, {Omodei},
  {Orienti}, {Orlando}, {Ormes}, {Paneque}, {Panetta}, {Perkins},
  {Pesce-Rollins}, {Piron}, {Pivato}, {Porter}, {Racusin}, {Rando}, {Razzano},
  {Razzaque}, {Reimer}, {Reimer}, {Reposeur}, {Rochester}, {Romani},
  {Salvetti}, {S{\'a}nchez-Conde}, {Saz Parkinson}, {Schulz}, {Siskind},
  {Smith}, {Spada}, {Spandre}, {Spinelli}, {Stephens}, {Strong}, {Suson},
  {Takahashi}, {Takahashi}, {Tanaka}, {Thayer}, {Thayer}, {Thompson},
  {Tibaldo}, {Tibolla}, {Torres}, {Torresi}, {Tosti}, {Troja}, {Van Klaveren},
  {Vianello}, {Winer}, {Wood}, {Wood}, {Zimmer}, \& {Fermi-LAT
  Collaboration}}]{2015ApJS..218...23A}
{Acero}, F., {Ackermann}, M., {Ajello}, M., {et~al.} 2015,
  \href{http://dx.doi.org/10.1088/0067-0049/218/2/23}{\JournalTitle{\apjs},
  218, 23}

\bibitem[{{Becker} {et~al.}(1995){Becker}, {White}, \&
  {Helfand}}]{1995ApJ...450..559B}
{Becker}, R.~H., {White}, R.~L., \& {Helfand}, D.~J. 1995,
  \href{http://dx.doi.org/10.1086/176166}{\JournalTitle{\apj}, 450, 559}

\bibitem[{{Bednarek} \& {Pabich}(2011)}]{2011A&A...530A..49B}
{Bednarek}, W., \& {Pabich}, J. 2011,
  \href{http://dx.doi.org/10.1051/0004-6361/201116549}{\JournalTitle{\aap},
  530, A49}

\bibitem[{{Benaglia} {et~al.}(2010){Benaglia}, {Romero}, {Mart{\'{\i}}},
  {Peri}, \& {Araudo}}]{2010A&A...517L..10B}
{Benaglia}, P., {Romero}, G.~E., {Mart{\'{\i}}}, J., {Peri}, C.~S., \&
  {Araudo}, A.~T. 2010,
  \href{http://dx.doi.org/10.1051/0004-6361/201015232}{\JournalTitle{\aap},
  517, L10}

\bibitem[{{Berezinskii} {et~al.}(1990){Berezinskii}, {Bulanov}, {Dogiel}, \&
  {Ptuskin}}]{1990acr..book.....B}
{Berezinskii}, V.~S., {Bulanov}, S.~V., {Dogiel}, V.~A., \& {Ptuskin}, V.~S.
  1990, {Astrophysics of cosmic rays}

\bibitem[{{Brose} {et~al.}(2016){Brose}, {Telezhinsky}, \&
  {Pohl}}]{2016A&A...593A..20B}
{Brose}, R., {Telezhinsky}, I., \& {Pohl}, M. 2016,
  \href{http://dx.doi.org/10.1051/0004-6361/201527345}{\JournalTitle{\aap},
  593, A20}

\bibitem[{{Canto} {et~al.}(1996){Canto}, {Raga}, \&
  {Wilkin}}]{1996ApJ...469..729C}
{Canto}, J., {Raga}, A.~C., \& {Wilkin}, F.~P. 1996,
  \href{http://dx.doi.org/10.1086/177820}{\JournalTitle{\apj}, 469, 729}

\bibitem[{{Caprioli} \& {Spitkovsky}(2014)}]{2014ApJ...783...91C}
{Caprioli}, D., \& {Spitkovsky}, A. 2014,
  \href{http://dx.doi.org/10.1088/0004-637X/783/2/91}{\JournalTitle{\apj}, 783,
  91}

\bibitem[{{Comeron} \& {Kaper}(1998)}]{1998A&A...338..273C}
{Comeron}, F., \& {Kaper}, L. 1998, \JournalTitle{\aap}, 338, 273

\bibitem[{{Condon} {et~al.}(1998){Condon}, {Cotton}, {Greisen}, {Yin},
  {Perley}, {Taylor}, \& {Broderick}}]{1998AJ....115.1693C}
{Condon}, J.~J., {Cotton}, W.~D., {Greisen}, E.~W., {et~al.} 1998,
  \href{http://dx.doi.org/10.1086/300337}{\JournalTitle{\aj}, 115, 1693}

\bibitem[{{De Becker} {et~al.}(2017){De Becker}, {del Valle}, {Romero}, {Peri},
  \& {Benaglia}}]{2017MNRAS.471.4452D}
{De Becker}, M., {del Valle}, M.~V., {Romero}, G.~E., {Peri}, C.~S., \&
  {Benaglia}, P. 2017,
  \href{http://dx.doi.org/10.1093/mnras/stx1826}{\JournalTitle{\mnras}, 471,
  4452}

\bibitem[{{de la Cita} {et~al.}(2016){de la Cita}, {Bosch-Ramon},
  {Paredes-Fortuny}, {Khangulyan}, \& {Perucho}}]{2016A&A...591A..15D}
{de la Cita}, V.~M., {Bosch-Ramon}, V., {Paredes-Fortuny}, X., {Khangulyan},
  D., \& {Perucho}, M. 2016,
  \href{http://dx.doi.org/10.1051/0004-6361/201527084}{\JournalTitle{\aap},
  591, A15}

\bibitem[{{del Palacio} {et~al.}(2018){del Palacio}, {Bosch-Ramon},
  {M{\"u}ller}, \& {Romero}}]{2018arXiv180610863D}
{del Palacio}, S., {Bosch-Ramon}, V., {M{\"u}ller}, A.~L., \& {Romero}, G.~E.
  2018, \JournalTitle{ArXiv e-prints},
  \href{http://arxiv.org/abs/1806.10863}{{\sffamily arXiv:1806.10863
  [astro-ph.HE]}}

\bibitem[{{del Valle} {et~al.}(2018){del Valle}, {M{\"u}ller}, \&
  {Romero}}]{2017arXiv171106250D}
{del Valle}, M.~V., {M{\"u}ller}, A.~L., \& {Romero}, G.~E. 2018,
  \href{http://dx.doi.org/10.1093/mnras/stx2984}{\JournalTitle{\mnras}, 475,
  4298}

\bibitem[{{del Valle} \& {Romero}(2012)}]{2012A&A...543A..56D}
{del Valle}, M.~V., \& {Romero}, G.~E. 2012,
  \href{http://dx.doi.org/10.1051/0004-6361/201218937}{\JournalTitle{\aap},
  543, A56}

\bibitem[{{del Valle} \& {Romero}(2014)}]{2014A&A...563A..96D}
---. 2014,
  \href{http://dx.doi.org/10.1051/0004-6361/201322308}{\JournalTitle{\aap},
  563, A96}

\bibitem[{{del Valle} {et~al.}(2013){del Valle}, {Romero}, \& {De
  Becker}}]{2013A&A...550A.112D}
{del Valle}, M.~V., {Romero}, G.~E., \& {De Becker}, M. 2013,
  \href{http://dx.doi.org/10.1051/0004-6361/201220112}{\JournalTitle{\aap},
  550, A112}

\bibitem[{{del Valle} {et~al.}(2015){del Valle}, {Romero}, \&
  {Santos-Lima}}]{2015MNRAS.448..207D}
{del Valle}, M.~V., {Romero}, G.~E., \& {Santos-Lima}, R. 2015,
  \href{http://dx.doi.org/10.1093/mnras/stu2732}{\JournalTitle{\mnras}, 448,
  207}

\bibitem[{{Dgani} {et~al.}(1996){Dgani}, {van Buren}, \&
  {Noriega-Crespo}}]{1996ApJ...461..927D}
{Dgani}, R., {van Buren}, D., \& {Noriega-Crespo}, A. 1996,
  \href{http://dx.doi.org/10.1086/177114}{\JournalTitle{\apj}, 461, 927}

\bibitem[{{Draine}(2011)}]{2011piim.book.....D}
{Draine}, B.~T. 2011, {Physics of the Interstellar and Intergalactic Medium}

\bibitem[{{Draine} \& {Li}(2007)}]{2007ApJ...657..810D}
{Draine}, B.~T., \& {Li}, A. 2007,
  \href{http://dx.doi.org/10.1086/511055}{\JournalTitle{\apj}, 657, 810}

\bibitem[{{Drury}(1983)}]{1983RPPh...46..973D}
{Drury}, L.~O. 1983,
  \href{http://dx.doi.org/10.1088/0034-4885/46/8/002}{\JournalTitle{Reports on
  Progress in Physics}, 46, 973}

\bibitem[{{Ellison} {et~al.}(1990){Ellison}, {Moebius}, \&
  {Paschmann}}]{1990ApJ...352..376E}
{Ellison}, D.~C., {Moebius}, E., \& {Paschmann}, G. 1990,
  \href{http://dx.doi.org/10.1086/168544}{\JournalTitle{\apj}, 352, 376}

\bibitem[{{Gaisser}(1990)}]{1990cup..book.....G}
{Gaisser}, T.~K. 1990, {Cosmic rays and particle physics}

\bibitem[{{H.~E.~S.~S.~Collaboration}
  {et~al.}(2017){H.~E.~S.~S.~Collaboration}, {:}, {Abdalla}, {Abramowski},
  {Aharonian}, {Ait Benkhali}, {Akhperjanian}, {Andersson}, {Ang{\"u}ner},
  {Arakawa}, \& et~al.}]{2017arXiv170502263H}
{H.~E.~S.~S.~Collaboration}, {:}, {Abdalla}, H., {et~al.} 2017,
  \JournalTitle{ArXiv e-prints},
  \href{http://arxiv.org/abs/1705.02263}{{\sffamily arXiv:1705.02263
  [astro-ph.HE]}}

\bibitem[{{Hoogerwerf} {et~al.}(2000){Hoogerwerf}, {de Bruijne}, \& {de
  Zeeuw}}]{2000ApJ...544L.133H}
{Hoogerwerf}, R., {de Bruijne}, J.~H.~J., \& {de Zeeuw}, P.~T. 2000,
  \href{http://dx.doi.org/10.1086/317315}{\JournalTitle{\apjl}, 544, L133}

\bibitem[{{Kobulnicky} {et~al.}(2010){Kobulnicky}, {Gilbert}, \&
  {Kiminki}}]{2010ApJ...710..549K}
{Kobulnicky}, H.~A., {Gilbert}, I.~J., \& {Kiminki}, D.~C. 2010,
  \href{http://dx.doi.org/10.1088/0004-637X/710/1/549}{\JournalTitle{\apj},
  710, 549}

\bibitem[{{Kobulnicky} {et~al.}(2016){Kobulnicky}, {Chick}, {Schurhammer},
  {Andrews}, {Povich}, {Munari}, {Olivier}, {Sorber}, {Wernke}, {Dale}, \&
  {Dixon}}]{2016ApJS..227...18K}
{Kobulnicky}, H.~A., {Chick}, W.~T., {Schurhammer}, D.~P., {et~al.} 2016,
  \href{http://dx.doi.org/10.3847/0067-0049/227/2/18}{\JournalTitle{\apjs},
  227, 18}

\bibitem[{{Lequeux}(2005)}]{2005ism..book.....L}
{Lequeux}, J. 2005, {The Interstellar Medium}

\bibitem[{{Longair}(2011)}]{2011hea..book.....L}
{Longair}, M.~S. 2011, {High Energy Astrophysics}

\bibitem[{{L{\'o}pez-Santiago} {et~al.}(2012){L{\'o}pez-Santiago}, {Miceli},
  {del Valle}, {Romero}, {Bonito}, {Albacete-Colombo}, {Pereira}, {de Castro},
  \& {Damiani}}]{2012ApJ...757L...6L}
{L{\'o}pez-Santiago}, J., {Miceli}, M., {del Valle}, M.~V., {et~al.} 2012,
  \href{http://dx.doi.org/10.1088/2041-8205/757/1/L6}{\JournalTitle{\apjl},
  757, L6}

\bibitem[{{Mac Low} \& {Norman}(1993)}]{1993ApJ...407..207M}
{Mac Low}, M.-M., \& {Norman}, M.~L. 1993,
  \href{http://dx.doi.org/10.1086/172506}{\JournalTitle{\apj}, 407, 207}

\bibitem[{{Meyer} {et~al.}(2014){Meyer}, {Mackey}, {Langer}, {Gvaramadze},
  {Mignone}, {Izzard}, \& {Kaper}}]{2014MNRAS.444.2754M}
{Meyer}, D.~M.-A., {Mackey}, J., {Langer}, N., {et~al.} 2014,
  \href{http://dx.doi.org/10.1093/mnras/stu1629}{\JournalTitle{\mnras}, 444,
  2754}

\bibitem[{{Meyer} {et~al.}(2017){Meyer}, {Mignone}, {Kuiper}, {Raga}, \&
  {Kley}}]{2017MNRAS.464.3229M}
{Meyer}, D.~M.-A., {Mignone}, A., {Kuiper}, R., {Raga}, A.~C., \& {Kley}, W.
  2017, \href{http://dx.doi.org/10.1093/mnras/stw2537}{\JournalTitle{\mnras},
  464, 3229}

\bibitem[{{Meyer} {et~al.}(2016){Meyer}, {van Marle}, {Kuiper}, \&
  {Kley}}]{2016MNRAS.459.1146M}
{Meyer}, D.~M.-A., {van Marle}, A.-J., {Kuiper}, R., \& {Kley}, W. 2016,
  \href{http://dx.doi.org/10.1093/mnras/stw651}{\JournalTitle{\mnras}, 459,
  1146}

\bibitem[{{Mignone} {et~al.}(2007){Mignone}, {Bodo}, {Massaglia}, {Matsakos},
  {Tesileanu}, {Zanni}, \& {Ferrari}}]{2007ApJS..170..228M}
{Mignone}, A., {Bodo}, G., {Massaglia}, S., {et~al.} 2007,
  \href{http://dx.doi.org/10.1086/513316}{\JournalTitle{\apjs}, 170, 228}

\bibitem[{{Mokiem} {et~al.}(2007){Mokiem}, {de Koter}, {Vink}, {Puls}, {Evans},
  {Smartt}, {Crowther}, {Herrero}, {Langer}, {Lennon}, {Najarro}, \&
  {Villamariz}}]{2007A&A...473..603M}
{Mokiem}, M.~R., {de Koter}, A., {Vink}, J.~S., {et~al.} 2007,
  \href{http://dx.doi.org/10.1051/0004-6361:20077545}{\JournalTitle{\aap}, 473,
  603}

\bibitem[{{Nolan} {et~al.}(2012){Nolan}, {Abdo}, {Ackermann}, {Ajello},
  {Allafort}, {Antolini}, {Atwood}, {Axelsson}, {Baldini}, {Ballet}, \&
  et~al.}]{2012ApJS..199...31N}
{Nolan}, P.~L., {Abdo}, A.~A., {Ackermann}, M., {et~al.} 2012,
  \href{http://dx.doi.org/10.1088/0067-0049/199/2/31}{\JournalTitle{\apjs},
  199, 31}

\bibitem[{{Pacholczyk}(1970)}]{1970ranp.book.....P}
{Pacholczyk}, A.~G. 1970, {Radio astrophysics. Nonthermal processes in galactic
  and extragalactic sources}

\bibitem[{{Pakmor} {et~al.}(2016){Pakmor}, {Pfrommer}, {Simpson}, \&
  {Springel}}]{2016ApJ...824L..30P}
{Pakmor}, R., {Pfrommer}, C., {Simpson}, C.~M., \& {Springel}, V. 2016,
  \href{http://dx.doi.org/10.3847/2041-8205/824/2/L30}{\JournalTitle{\apjl},
  824, L30}

\bibitem[{{Pereira} {et~al.}(2016){Pereira}, {L{\'o}pez-Santiago}, {Miceli},
  {Bonito}, \& {de Castro}}]{2016A&A...588A..36P}
{Pereira}, V., {L{\'o}pez-Santiago}, J., {Miceli}, M., {Bonito}, R., \& {de
  Castro}, E. 2016,
  \href{http://dx.doi.org/10.1051/0004-6361/201527985}{\JournalTitle{\aap},
  588, A36}

\bibitem[{{Peri} {et~al.}(2012){Peri}, {Benaglia}, {Brookes}, {Stevens}, \&
  {Isequilla}}]{2012A&A...538A.108P}
{Peri}, C.~S., {Benaglia}, P., {Brookes}, D.~P., {Stevens}, I.~R., \&
  {Isequilla}, N.~L. 2012,
  \href{http://dx.doi.org/10.1051/0004-6361/201118116}{\JournalTitle{\aap},
  538, A108}

\bibitem[{{Peri} {et~al.}(2015){Peri}, {Benaglia}, \&
  {Isequilla}}]{2015A&A...578A..45P}
{Peri}, C.~S., {Benaglia}, P., \& {Isequilla}, N.~L. 2015,
  \href{http://dx.doi.org/10.1051/0004-6361/201424676}{\JournalTitle{\aap},
  578, A45}

\bibitem[{{Pohl}(1993)}]{1993A&A...270...91P}
{Pohl}, M. 1993, \JournalTitle{\aap}, 270, 91

\bibitem[{{Raga} {et~al.}(1997){Raga}, {Noriega-Crespo}, {Cant{\'o}},
  {Steffen}, {van Buren}, {Mellema}, \& {Lundqvist}}]{1997RMxAA..33...73R}
{Raga}, A.~C., {Noriega-Crespo}, A., {Cant{\'o}}, J., {et~al.} 1997,
  \JournalTitle{\rmxaa}, 33, 73

\bibitem[{{Romero} {et~al.}(2010){Romero}, {Del Valle}, \&
  {Orellana}}]{2010A&A...518A..12R}
{Romero}, G.~E., {Del Valle}, M.~V., \& {Orellana}, M. 2010,
  \href{http://dx.doi.org/10.1051/0004-6361/200913938}{\JournalTitle{\aap},
  518, A12}

\bibitem[{{Ryu} \& {Vishniac}(1987)}]{1987ApJ...313..820R}
{Ryu}, D., \& {Vishniac}, E.~T. 1987,
  \href{http://dx.doi.org/10.1086/165021}{\JournalTitle{\apj}, 313, 820}

\bibitem[{{S{\'a}nchez-Ayaso} {et~al.}(2018){S{\'a}nchez-Ayaso}, {del Valle},
  {Mart{\'{\i}}}, {Romero}, \& {Luque-Escamilla}}]{2018arXiv180600614S}
{S{\'a}nchez-Ayaso}, E., {del Valle}, M.~V., {Mart{\'{\i}}}, J., {Romero},
  G.~E., \& {Luque-Escamilla}, P.~L. 2018, \JournalTitle{ArXiv e-prints},
  \href{http://arxiv.org/abs/1806.00614}{{\sffamily arXiv:1806.00614
  [astro-ph.HE]}}

\bibitem[{{Schulz} {et~al.}(2014){Schulz}, {Ackermann}, {Buehler}, {Mayer}, \&
  {Klepser}}]{2014A&A...565A..95S}
{Schulz}, A., {Ackermann}, M., {Buehler}, R., {Mayer}, M., \& {Klepser}, S.
  2014,
  \href{http://dx.doi.org/10.1051/0004-6361/201423468}{\JournalTitle{\aap},
  565, A95}

\bibitem[{{Telezhinsky} {et~al.}(2012){Telezhinsky}, {Dwarkadas}, \&
  {Pohl}}]{2012A&A...541A.153T}
{Telezhinsky}, I., {Dwarkadas}, V.~V., \& {Pohl}, M. 2012,
  \href{http://dx.doi.org/10.1051/0004-6361/201118639}{\JournalTitle{\aap},
  541, A153}

\bibitem[{{Tetzlaff} {et~al.}(2011){Tetzlaff}, {Neuh{\"a}user}, \&
  {Hohle}}]{2011MNRAS.410..190T}
{Tetzlaff}, N., {Neuh{\"a}user}, R., \& {Hohle}, M.~M. 2011,
  \href{http://dx.doi.org/10.1111/j.1365-2966.2010.17434.x}{\JournalTitle{\mnras},
  410, 190}

\bibitem[{{Toal{\'a}} {et~al.}(2016){Toal{\'a}}, {Oskinova},
  {Gonz{\'a}lez-Gal{\'a}n}, {Guerrero}, {Ignace}, \&
  {Pohl}}]{2016ApJ...821...79T}
{Toal{\'a}}, J.~A., {Oskinova}, L.~M., {Gonz{\'a}lez-Gal{\'a}n}, A., {et~al.}
  2016,
  \href{http://dx.doi.org/10.3847/0004-637X/821/2/79}{\JournalTitle{\apj}, 821,
  79}

\bibitem[{{Toal{\'a}} {et~al.}(2017){Toal{\'a}}, {Oskinova}, \&
  {Ignace}}]{2017ApJ...838L..19T}
{Toal{\'a}}, J.~A., {Oskinova}, L.~M., \& {Ignace}, R. 2017,
  \href{http://dx.doi.org/10.3847/2041-8213/aa667c}{\JournalTitle{\apjl}, 838,
  L19}

\bibitem[{{van Buren} \& {McCray}(1988)}]{1988ApJ...329L..93V}
{van Buren}, D., \& {McCray}, R. 1988,
  \href{http://dx.doi.org/10.1086/185184}{\JournalTitle{\apjl}, 329, L93}

\bibitem[{{van Marle} {et~al.}(2011){van Marle}, {Meliani}, {Keppens}, \&
  {Decin}}]{2011ApJ...734L..26V}
{van Marle}, A.~J., {Meliani}, Z., {Keppens}, R., \& {Decin}, L. 2011,
  \href{http://dx.doi.org/10.1088/2041-8205/734/2/L26}{\JournalTitle{\apjl},
  734, L26}

\bibitem[{{Voelk} \& {Forman}(1982)}]{1982ApJ...253..188V}
{Voelk}, H.~J., \& {Forman}, M. 1982,
  \href{http://dx.doi.org/10.1086/159623}{\JournalTitle{\apj}, 253, 188}

\bibitem[{{Walder} {et~al.}(2012){Walder}, {Folini}, \&
  {Meynet}}]{2012SSRv..166..145W}
{Walder}, R., {Folini}, D., \& {Meynet}, G. 2012,
  \href{http://dx.doi.org/10.1007/s11214-011-9771-2}{\JournalTitle{\ssr}, 166,
  145}

\bibitem[{{Wiersma} {et~al.}(2009){Wiersma}, {Schaye}, \&
  {Smith}}]{2009MNRAS.393...99W}
{Wiersma}, R.~P.~C., {Schaye}, J., \& {Smith}, B.~D. 2009,
  \href{http://dx.doi.org/10.1111/j.1365-2966.2008.14191.x}{\JournalTitle{\mnras},
  393, 99}

\bibitem[{{Wilkin}(1996)}]{1996ApJ...459L..31W}
{Wilkin}, F.~P. 1996,
  \href{http://dx.doi.org/10.1086/309939}{\JournalTitle{\apjl}, 459, L31}

\bibitem[{{Wilkin}(2000)}]{2000ApJ...532..400W}
---. 2000, \href{http://dx.doi.org/10.1086/308576}{\JournalTitle{\apj}, 532,
  400}

\end{thebibliography}

\listofchanges 
\end{document}